\documentstyle[psfig,epsf]{ar}

\begin{document}

\newcommand{\beq}{\begin{equation}}
\newcommand{\eeq}{\end{equation}}
\newcommand{\beqa}{\begin{eqnarray}}
\newcommand{\eeqa}{\end{eqnarray}}
\newcommand{\boldsigma}{\mbox{\boldmath$\sigma$}}
\newcommand{\boldtau}{\mbox{\boldmath$\tau$}}
\newcommand{\ott}{\boldtau_i \cdot \boldtau_j}
\newcommand{\oss}{\boldsigma_i\cdot\boldsigma_j}
\newcommand{\os}{S_{ij}}
\newcommand{\ols}{({\bf L\cdot S})_{ij}}
\newcommand{\hovm}{\frac{\hbar^2}{m}}
\newcommand{\hovtm}{\frac{\hbar^2}{2m}}
\newcommand \beqas{\begin{eqnarray*}}
\newcommand \eeqas{\end{eqnarray*}}
\newcommand {\veb}[1]{\bf{#1}}
\newcommand \la{\raisebox{-.5ex}{$\stackrel{<}{\sim}$}}
\newcommand \ga{\raisebox{-.5ex}{$\stackrel{>}{\sim}$}}
\newcommand \degs{^{\circ}}
\newcommand \nuc[2]{$^{#2}$#1}
\newcommand {\msol}{$M_\odot \;$}


\title{Recent Progress in Neutron Star Theory}


\author{Henning Heiselberg
       \affiliation{NORDITA, Blegdamsvej 17, DK-2100 Copenhagen \O, Denmark}
       Vijay Pandharipande
       \affiliation{Department of Physics, University of Illinois at
Urbana-Champaign, 1110 W. Green St., Urbana, Illinois 61801, USA}}

\date{\today}
\maketitle
\begin{abstract}

\end{abstract}


\section{Introduction}

Neutron stars are among the most fascinating bodies in our universe. 
They contain over a solar mass of matter within a radius of $\sim$ 10 km 
at densities of order $10^{15}$ g/cc.  They probe the properties of cold 
matter at extremely high densities, and 
have proven to be fantastic test bodies for theories of  
general relativity.  In a broader
perspective, neutron stars and heavy ion collisions provide access 
to the phase diagram of matter at extreme densities and temperatures, 
that is basic for understanding the very early Universe and several other 
astrophysical phenomena.

The discovery of the neutron by Chadwick  
in 1932 prompted Landau \cite{rosenfeld} to
predict the existence of neutron stars. 
The first theoretical calculations of neutron stars were performed by 
Oppenheimer and Volkoff \cite{Oppe39} in 1939 assuming that they are 
gravitationally bound states of neutron Fermi gas. 
The calculated stars had 
a maximum mass of $\sim$ 0.7 $M_\odot$, central 
densities up to $\sim 6 \times 10^{15}$~g/cm$^3$ and
radii $\sim10$~km.  For comparison 
the density of nuclear matter inside a large nucleus like $^{208}$Pb 
is $\sim$ 0.16 nucleons/fm$^3$,
{\em i.e.}  $\simeq$ 2.7 $\times 10^{14}$~g/cm$^3$ \cite{BohrMo}. 
Their predicted maximum mass was less than the 
Chandrasekhar mass limit of $\sim$ 1.4 $M_\odot$ for white 
dwarfs made up of iron group nuclei, and having densities up to 
$\sim 10^9$~g/cm$^3$ \cite{ShapTu}.  
The pressure to balance the gravitational attraction in white 
dwarfs and Oppenheimer-Volkoff neutron stars is 
supplied by degenerate electron and neutron Fermi gases 
respectively. 

In 1934 Baade and Zwicky \cite{Baad34} suggested that neutron
stars may be formed in {\it supernovae} in which the iron core of a
massive star exceeds the Chandrasekhar limit and collapses.  The large
amount of energy released in the collapse blows away the rest of the
star and the collapsed core may form a neutron star.  For efficient
production of neutron stars with this mechanism, the maximum mass of
neutron stars should exceed 1.4 $M_\odot$.  In the 60's, using
schematic models of nuclear forces, Tsuruta and Cameron \cite{camsol}
showed that they could increase the neutron star masses beyond 1.4
$M_\odot$.

Bell and Hewish discovered {\it radio pulsars} in 1967, and they were
soon identified as rotating neutron stars by Gold \cite{Gold69}.  The
subsequent detection of the Crab pulsar in the remnant of the Crab
supernova, observed in China in 1054 A.D., confirmed the link to
supernovae, and initiated the present efforts to better understand 
neutron stars.  

\subsection{A Brief Overview of Observations} 

Almost 1200 pulsars have been discovered by the turn of this
millennium.  In these stars the magnetic and rotational axes are
misaligned, thus they emit dipole radiation in the form of radio waves
that appear to pulse on and off like a lighthouse beacon as the pulsar
beam sweeps across the Earth.  The rotational energy loss due to
dipole radiation is
\beq
 \dot{E} = I\Omega\dot{\Omega} 
         = - \frac{B^2R^6\Omega^4\sin^2\theta}{6c^3} \,, \label{dE}
\eeq
where the moment of inertia for a typical neutron star is $I\sim
10^{45}$~g~cm$^2$.  Pulsars have magnetic fields $B$ of $\sim 10^{12}$
G, deduced from the observed $\dot{\Omega}$, and independently
confirmed by cyclotron absorption lines found in X-ray spectra.  Their
periods, $P = 2 \pi / \Omega$, ranging from 1.5~ms to 8.5~s, are
increasing with derivatives $\dot{P}\sim 10^{-12}-10^{-21}$.  The
pulsar age is approximately given by $ P/2\dot{P}$ \cite{Lorimer};
most pulsars are old and slowly rotating with relatively small period
derivatives, except for a few young pulsars, e.g., those found in the
Crab and Vela nebulae.

In 1969 the Crab and the Vela pulsars were observed to {\it "glitch"},
{\em i.e.} to suddenly speedup with period changes $\Delta P/P$ of the
order of $10^{-8}$ and $10^{-6}$ respectively \cite{Boyn72}.  In
post-glitch relaxation most of the period increase $\Delta P$ decays.
These pulsars have glitched several times since then.  The glitches
suggest that the neutron stars have a solid crust containing
superfluid neutrons.  The interesting structure of their crust has
been recently reviewed \cite{prannr}, and we discuss it rather briefly
in this report.

The first {\it binary} of two {\it pulsars} was found by Hulse and
Taylor in 1973 and they could determine many of its parameters
including both masses, orbital period and period derivative, orbital
distance and inclination. General relativity could be tested to an
unprecedented accuracy by measuring the inward spiralling of the
neutron stars in the Hulse-Taylor binary PSR 1913+16
\cite{HulseT}. The periastron
advance in PSR 1913+16 is 4.2$^\circ$ per year as compared to 43'' per
century for Mercury, which originally was used by Einstein to test his
theory of general relativity.  Six double neutron star binaries are
known so far, and neutron stars in all of them have masses in the
range $1.36\pm 0.08 M_\odot$ \cite{Thor99}.  They 
confirm that nuclear forces have a large effect on the structure of
neutron stars and increase their maximum mass beyond 1.4 $M_\odot$.
Neutron stars are estimated to have a binding energy of $\sim 10 \%$
of their mass.  Thus $\sim 1.5$ $M_\odot$ of nuclei are needed to
obtain a $1.35$ $M_\odot$ star.

A distinct subclass of radio pulsars are {\it millisecond pulsars}
with periods $\la 100$ ms. The fastest pulsar known has a period of
1.56~ms \cite{fastes}.  
The period derivatives of millisecond pulsars are very small
corresponding to low magnetic fields $\sim 10^8-10^{10}$G.  They are
believed to be recycled pulsars, i.e.\ old pulsars that have been spun
up by mass accretion whereby the magnetic fields have decayed.
About 80\% of the millisecond pulsars are in binaries whereas less than
1\% of normal radio pulsars are in binaries.  About 20 - almost half
of the millisecond pulsars - are found in binaries where the companion
is either a white dwarf or a neutron star.

With X-ray detectors on board satellites since the early 1970's about
two hundred {\it X-ray pulsars and bursters} have been found of which
the rotational period has been determined for about sixty.  The X-ray
pulsars and bursters \cite{BildStroh} are believed to be neutron stars
accreting matter from high ($M\ga 10M_\odot$) and low mass ($M\la
1.2M_\odot$) companions respectively. The X-ray pulses are attributed
to strong accretion on the magnetic poles emitting X-rays (as northern
lights).  The observed radiation is pulsed with the rotational
frequency of the accreting star.  
{\em X-ray bursts} are thermonuclear explosions of accreted matter on
the surface of neutron stars. After accumulating hydrogen on the
surface for hours, pressure and temperature become sufficient to
trigger a runaway thermonuclear explosion seen as an X-ray burst that
lasts a few seconds \cite{BildStroh}.
Masses of these stars are less accurately measured than for
binary pulsars.  We mention recent mass determinations for the X-ray
pulsar Vela X-1: $M=1.87^{+0.23}_{-0.17} \ M_\odot$ \cite{Barziv},  
and the burster Cygnus X-2: $M=1.8\pm0.4)M_\odot$
\cite{Orosz}. They are larger than the typical $1.36 \pm 0.08 M_\odot$ masses
found in pulsars binaries, presumably due to accreted matter.

A subclass of half a dozen {\em anomalous X-ray pulsars} has been
discovered. They are slowly rotating, $P\sim 10$~sec, but rapidly
slowing down. This requires huge magnetic fields of $B\sim10^{14}$~G
and they have appropriately been named ``magnetars'' \cite{Duncan}.
Four gamma ray repeaters discovered so far are also believed to be
slowly rotating neutron stars. The magnetars and likely also the
gamma ray repeaters reside inside supernova remnants.

Recently, {\em quasi-periodic oscillations} (QPO) have been found in
12 binaries of neutron stars with low mass companions.  If the QPO
originate from the innermost stable orbit \cite{zss97,miller} of the
accreting matter, their observed values imply that the accreting
neutron stars have masses up to $\simeq 2.3M_\odot$.  In this case
the QPO's also constrain the radii of the accreting star.

Non-rotating and non-accreting neutron stars are virtually
undetectable but the Hubble space telescope has observed one thermally
radiating neutron star \cite{Walter}.  Its surface temperature is
$T\simeq 6\times 10^5$~K$\simeq 50$~eV and its distance is less than
120~pc from Earth. Circumstantial evidence indicate a distance of
$\sim80$~pc which leads to a radius of 12-13~km for this star.  In
recent years much effort has been devoted to measuring pulsar
temperatures, especially with the Einstein Observatory and ROSAT.
Surface temperatures of a few pulsars have been measured, and upper
limits have been set for many \cite{pethick}.

>From the human point of view supernova explosions are  
rare in our and neighboring galaxies.  The predicted rate is 1-3 per
century in our galaxy and the most recent one was 1987A in LMC.  No
neutron star associated with this explosion has been detected; 
however, 19 neutrinos were detected on earth from 1987A \cite{1987an}, 
indicating the formation of a ``proto-neutron star''.  
It has been suggested by Bethe and Brown \cite{Bebhole} that an 
upper limit to the mass of neutron stars can be obtained assuming that 
the remnant of SN 1987A collapsed into a black hole. 

Astrophysicists expect a large abundance of $\sim 10^8$ neutron
stars in our galaxy.  At least as many supernova explosions, responsible for all
heavier elements present in our Universe today, have occurred.  The
scarcity of neutron stars in the solar neighborhood may be due to 
production of black holes or other remnants in supernovae, or due to a
high initial velocity (asymmetric ``kick'') received during their birth in
supernovae. Recently, many neutron stars have been found far away from
their supernova remnants; and of the $\sim1200$ discovered radio pulsars
only about $\sim10$ can be associated with the 220 known supernova
remnants.  Neutron stars thrown out of the galactic
plane may be detected by gravitational microlensing experiments
\cite{gmicro} designed to search for dark massive objects in the galactic
halo.

The recent discovery of afterglow in {\em Gamma Ray Bursters} (GRB)
allows determination of their very high redshifts ($z\ge 1$).  They
imply that GRB occur at enormous distances. Evidence for beaming has
been observed \cite{Kulkar}, and the estimated energy output is $\sim
10^{53}$ ergs. Such enormous energies can be produced in neutron star
mergers eventually forming black holes. From abundance of binary
pulsars one can estimate the rate of neutron star mergers; it is
compatible with the rate of GRB of approximately one per day.  Another
possible mechanism, is a special class of type Ic supernova ({\it hypernovae})
where cores collapse to black holes \cite{hypernovae}.

{\em The future} of neutron star observations looks bright as new windows are
about to open. A new fleet of X- and Gamma-ray satellites have and
will be launched. With upgraded ground based observatories and
detectors for neutrinos and gravitational waves \cite{LIGO}  
our knowledge of neutron star properties will be greatly improved.

\subsection{Theory of Neutron Star Matter}

Neutron stars are made up of relatively cold, charge neutral matter
with densities up to $\sim 7$ times the equilibrium density $\rho_0$ =
0.16 nucleons/fm$^3$ of charged nuclear matter in nuclei.  The matter
density is $> \rho_0$ over most of the star, apart from the relatively
thin crust \cite{prannr}. The Fermi energy of neutron star matter is in excess of
tens of MeV, and hence, at typical temperatures of $\la$ KeV, 
thermal effects are a minor perturbation on the gross structure of the star. 

Matter at such densities has not yet been produced in the laboratory, its 
properties must be theoretically deduced from the available terrestrial data
with guidance from observed neutron star properties. 
The quantities of interest are the phase and composition 
of cold catalyzed neutral dense matter, its energy  
density $\epsilon (\rho)$ 
and pressure $P(\rho)$, where $\rho$ denotes the baryon number density.  
The baryon number is conserved in all known 
interactions, therefore it is convenient to find the composition  
by minimizing the total energy $E_T(\rho)$ per baryon, including rest 
mass contributions.  This gives:
\beq
\epsilon(\rho)=\rho E_T(\rho),
\ \ \ \ P(\rho)= \rho^2 \frac{\partial E_T(\rho)} 
{\partial \rho}. 
\label{eq:eandp}
\eeq
The equation of state (EOS) $P(\epsilon)$ is found by eliminating 
$\rho$ from the above two.  

The gravitational equilibrium of a nonrotating star is described by the 
Tolman-Oppenheimer-Volkoff (TOV) \cite{ShapTu} Eq: 
\beq
\frac{dP(r)}{dr} = - \frac{G(\epsilon(r)+P(r)/c^2)(m(r)+4 \pi r^3 P(r)/c^2)} 
{r^2 (1-2Gm(r)/rc^2)} \ ,
\label{eq:tov}
\eeq
where $G$ is the gravitational constant, $P(r)$ and $\epsilon(r)$ are the 
pressure and mass density at radius $r$ in the star, and 
\beq
  m(r) = \int_0^r 4 \pi r^{\prime 2} \epsilon (r^{\prime}) dr^\prime \ , 
\label{eq:minr}
\eeq
is the mass inside $r$.  If we neglect the general relativistic corrections 
of order $1/c^2$ the TOV Eq. reduces to the Newtonian hydrodynamic equation. 
The TOV Eq. can be easily integrated starting 
from the central density $\epsilon_c$ ar $r=0$ to find the density profile 
$\epsilon(r)$.  At the radius $R$ of the star $P(R) = 0$, and $m(R) = M$ 
is the mass of the star as seen from outside.  The stability of the star 
can be deduced from the $M(\epsilon_c)$ as discussed in \cite{ShapTu}, 
and the equations for rotating stars are given by \cite{rotats}.
The effect of rotation on the structure of most observed neutron stars 
seems to be rather small, however, it could be significant at periods 
less than a millisecond \cite{cookstu}. 

At densities $< 2 \times 10^{-3} \rho_0 $  
matter is believed to have the form of a lattice of nuclei in a relativistic 
degenerate electron gas \cite{prannr}, qualitatively similar to 
that of metals.   
The main focus of the theory reviewed here has been on determining 
the properties and EOS of matter in the density 
range $2 \times 10^{-3} \rho_0 < \rho < 10 \rho_0$ from terrestrial data.  
In the lower part of this range we 
expect to find nucleon matter (NM) composed of nucleons and electrons.  In 
contrast to matter in nuclei, it has mostly neutrons with a small 
fraction of protons and equal number of electrons to maintain 
charge neutrality.  The large Fermi energy, $\mu_e \sim$ 100 MeV,
of the electron gas limits the fraction of protons in NM.  

At higher densities there are several possibilities including 
condensation of negatively charged pions and kaons, occurrence of 
hyperons, and the transition from hadronic to quark matter.  
All these possibilities exploit the large electron Fermi energy 
of NM, therefore only one of these, if any, may occur 
and lower the $\mu_e$. In addition, neutron star matter 
can have interesting mixed phase regions in which  
the mixing phases are charged but the matter is overall neutral
\cite{prannr}. 

We begin with a review of NM, and later consider the 
more exotic possibilities.  In the last sections the range of 
neutron star structures predicted by theory is presented along 
with a comparison with the observational data. 

\section{Energy-Density Functionals of Nucleon Matter}

The simplest description of nuclei is obtained within the mean field 
approximation.  It assumes, following the nuclear shell model, that nucleons 
occupy single particle orbitals in an average potential well produced by 
nuclear forces.  The energy of the nucleus is assumed to be a functional 
of the orbitals occupied by the nucleons, and the
orbitals are determined variationally as in the Hartree-Fock approximation.
In reality the mean field approximation is not exactly valid for nuclei.
The observed proton knockout reaction rates \cite{pswrmp} indicate that the 
shell model orbitals are occupied with a probability of $\sim 70 \% $ 
in the simplest closed shell nuclei like $^{208}$Pb. The differences
between the real and the mean field wave-functions, due to 
correlations induced by nuclear forces, are subsumed in the
energy functional as suggested by Kohn \cite{kohned} in the context of
atomic and molecular physics. 

The energy density of hypothetical, uniform NM at zero
temperature is the main term in the energy functionals.  The nucleon
orbitals in uniform matter are simple plane waves, and the ground
state in mean field approximation 
is obtained by filling the proton and neutron states up to their
Fermi momenta $k_{F,N} = (3 \pi^2 \rho_N)^{1/3}$, where $N = n,p $ for
neutrons and protons.  The energy density, denoted by ${\cal E}(\rho_n
, \rho_p)$, includes kinetic and strong interaction contributions, but
excludes rest masses and the Coulomb interaction, which destabilizes
uniform charged matter. The total density is denoted by $\rho = \rho_n
+ \rho_p $, the asymmetry of the matter is defined as $\beta = (\rho_n
- \rho_p)/\rho $, and the energy per nucleon, $E(\rho , \beta)$, is
given by $ {\cal E}(\rho , \beta )/ \rho $.  

Analysis of nuclear properties with the liquid drop model
\cite{BohrMo} reveals that, in the absence of electromagnetic forces,
the ground state of NM is symmetric $( i.e.\ \beta=0) $,
has total equilibrium density $\rho_0 = 0.16 \pm 0.01 $ fm$^{-3}$, and
binding energy $E_0 = - 16 \pm 0.5$ MeV per nucleon.  The symmetry
energy $E_{sym}(\rho_0) = 34 \pm 6$ MeV, is defined as $\frac{1}{2}
\partial^2 E/\partial\beta^2$ at equilibrium.  The NM
energy $E(\rho,\beta)$ can be expanded about its minimum value at
$\beta = 0$ in powers of $\beta^2$, assuming charge symmetry of
nuclear forces.  In variational \cite{lagasm} as well as Brueckner
\cite{bombaci91} theories the coefficients of terms with $\beta^{n
\geq 4}$ are estimated to be small, and $E(\rho , \beta) \approx
(1-\beta^2) E(\rho ,0) + \beta^2 E(\rho ,1)$.  In this
approximation the symmetry energy is the difference between the energy
of pure neutron matter and symmetric nuclear matter.  The
incompressibility $K_0 = 240 \pm 30$ MeV \cite{incomp} of symmetric
nuclear matter is defined as $K_0 = k_F^2 \partial^2 E / \partial
k_F^2 $ at equilibrium.  The energies of the collective breathing mode
vibrations of nuclei are sensitive to $K_0$; however, in all stable
nuclei the surface effects are significant.  It is difficult to
extract the density and $\beta$-dependence of the incompressibility,
and the density-dependence of the symmetry energy from available
nuclear data.

Analysis of elastic scattering of nucleons by nuclei shows that the
nuclear mean field has a dependence on the energy of the moving
nucleon \cite{BohrMo}.  Over a wide energy range this dependence is
approximately linear, suggesting that nucleons in equilibrium nuclear
matter have an effective mass $ m^{\star} \sim 0.7 m$, where $m$ is
the free nucleon mass.  This effective mass should not be identified with
the Landau effective mass which describes the density of single
particle states in a narrow energy interval about the Fermi energy
\cite{ffpop,liege76}.  The Landau $m^{\star}$ in uniform matter is difficult
to extract from nuclear data, since nucleons at the Fermi energy are
strongly coupled to nuclear surface dynamics.  Some of the
phenomenological energy functionals are chosen to fit the observed
nuclear level densities at the Fermi energy, while others fit the
value of $m^{\star}(\rho_0, 0) $ obtained from the energy dependence
of the optical model potential \cite{qfskyr}.

The nonrelativistic functionals based on Skyrme effective interactions 
\cite{qfskyr} generally contain the following terms:
\beq
{\cal E}(\rho_n , \rho_p) = \tau (1 + x_5 \rho) + x_1 \rho^2 ( 1 + x_2 
\rho^{\alpha} ) + \sum_{N=n,p} \left[ x_6 \tau_N \rho_N + 
x_3 \rho_N^2 ( 1 + x_4 \rho^{\alpha} ) \right] .
\label{eq:skyed}
\eeq
Here $\tau_N = 0.6 k_{F,N}^2 \rho_N /m $ 
are the kinetic energy densities, and $\tau = \tau_n + \tau_p $.  The 
parameters $x_1$ to $x_4$ and $\alpha$ describe the $\rho$ and $\beta$ 
dependence of the volume integral of the static part of the effective
interaction between nucleons in matter, while the $x_5$ and $x_6$ 
describe effective masses produced by the momentum dependence of the 
effective interaction.  In principle the values of the 
seven parameters in a typical Skyrme functional are constrained by 
the empirically known values of $\rho_0 ,\ E_0, \ E_{sym}(\rho_0)$ and
$K $ and the choice made for $m^{\star}(\rho_0, 0)$.  However, since the 
constraints are insufficient, there are many Skyrme models of the 
energy functional.  

The simple form of the functional (Eq.\ref{eq:skyed}) chosen by most Skyrme 
models is convenient, but the real functional can be much more complex.   
The analytic form of the energy density predicted by 
realistic models of nuclear forces, as discussed in the next chapter, 
has been studied by Ravenhall \cite{prlesh}.  A much more elaborate 
function of the type:
\beqa
{\cal E}(\rho_n , \rho_p) = - \rho^2 \left[ p_1 e^{-p_6 \rho} + p_2 
(1 - e^{-p_6 \rho}) + \left(\frac{p_{10}}{\rho} + p_{11} \right) 
e^{(p_9 \rho )^2} \right] \nonumber   \\
- \frac{1}{4} (\rho_n - \rho_p)^2 \left[ p_7 e^{-p_6 \rho} + p_8 
(1 - e^{-p_6 \rho}) + \left(\frac{p_{12}}{\rho} + p_{13} \right) 
e^{(p_9 \rho )^2} \right] \nonumber   \\
+ \sum_{N=n,p} \tau_N \left[ 1 + (p_3 \rho + p_5 \rho_N ) e^{-p_4 \rho}
\right],
\label{eq:rsid}
\eeqa
is required to reproduce the predicted ${\cal E}(\rho_n , \rho_p)$ up to $\beta = 1$ 
and $\rho \sim 1$ fm$^{-3}$.  This functional also explains the 
nuclear binding energies and the empirically known values for symmetric 
nuclear matter \cite{rsqual}, however, 
it is unlikely that the values of all of its thirteen parameters can be 
obtained by fitting nuclear data. 

The energy of NM can be easily calculated from a covariant
effective Lagrangian in the mean field approximation, as shown by 
Walecka \cite{swadv}, and in the past decade many properties 
of medium and heavy nuclei have been studied with this approach
\cite{ringan}.  The effective Lagrangian used in the recent work 
has the form:
\beqa
{\cal L} = \bar{\psi} \left[ \gamma^{\mu} \left( i \partial_{\mu} 
-g_{\omega} \omega_{\mu} -g_{\rho} \vec{\tau} \cdot \vec{\rho}_{\mu} 
\right) -m -g_{\sigma} \sigma \right] \psi -\frac{1}{2} m_{\sigma}^2 
\sigma^2  +\frac{1}{3} g_2 \sigma^3 + \frac{1}{4} g_3 \sigma^4  \nonumber   \\
 + \frac{1}{2}m^2_{\omega}\omega^{\mu}\omega_{\mu} 
+\frac{1}{2}m^2_{\rho}\vec{\rho}^{\mu} \cdot \vec{\rho}_{\mu} 
-\frac{1}{4}\Omega^{\mu \nu} \Omega_{\mu \nu}
-\frac{1}{4}\vec{R}^{\mu \nu} \cdot \vec{R}_{\mu \nu}, 
\label{eq:effl}
\eeqa
Hear $\psi ,\ \omega_{\mu}$ and $ \vec{\rho}_{\mu}$ are 
respectively the nucleon and $\omega$ and $\rho$ vector-meson  
fields.  Overhead arrows are used to denote isospin 
vectors, and $\Omega^{\mu \nu} = \partial^{\mu} \omega^{\nu}
- \partial^{\nu} \omega^{\mu} $ etc.  The effective scalar field 
$\sigma$ is responsible for nuclear binding, and the $\sigma^3$ and
$\sigma^4$ terms are necessary to obtain the empirical incompressibility of 
nuclear matter \cite{bbnl}.  The isovector $\vec{\rho}$ field 
is required to obtain the empirical symmetry energy.  The observed 
values of the masses $m,\ m_{\omega}$ and $m_{\rho}$ are used, and 
the coupling constants $g_{\omega},\ g_{\rho},\ g_{\sigma},\ g_2 $ 
and $g_3$, as well as the mass $m_{\sigma}$ of the effective scalar field  
are adjusted to fit the nuclear data.  The above Lagrangian, without the 
$\sigma^3$ and $\sigma^4$ terms but including pion fields and 
their coupling to the nucleon, is also used to model the two-nucleon 
interaction discussed in the next section.

The relativistic mean field theory of nuclei is very elegant and 
often used to study properties of neutron star matter \cite{glenb}.
It has provided important insights into relativistic effects in 
nuclei and NM.  However, the effective mean-field Lagrangian 
(Eq.\ref{eq:effl}) is unlikely to have a simple physical meaning. 
The inverse masses of the vector and scalar fields correspond to lengths 
of $\sim$ 0.25 and 0.4 fm, which are much smaller than the unit 
radius $r_0 = (4 \pi \rho /3)^{-1/3} \sim 1.2$ fm for equilibrium 
nuclear matter.  The naive 
condition for the validity of the mean field approximation, 
that $r_0$ be much less than the inverse masses of the fields is 
totally violated in nuclei as well as in neutron stars.  Pions 
are omitted from the effective Lagrangian because they do not 
contribute to the energy of matter in the mean field approximation. 
Their higher order contributions are subsumed in the effective 
scalar field.  Therefore, the 
effective mean-field Lagrangian must be interpreted as a relativistic 
generalization of Kohn's energy functional.  Eq.~(\ref{eq:effl}) 
assumes the simplest form necessary to fit nuclear data.  A more 
general form, necessary to explain the properties of NM 
over the wide density-asymmetry range in neutron stars, can have 
additional fields, isovector scalar for example, density dependent 
coupling constants to take into account 
the changes in correlations with density, and field energies  
containing  high powers of the fields, etc.  

The energy of low density neutron matter is well determined by
realistic models of two-nucleon interaction obtained by fitting the
nucleon-nucleon (NN) scattering data.  Different models and methods of
calculation give very similar results up to $\rho \sim \rho_0$, beyond
which three-nucleon interactions and relativistic effects, as well as
computational difficulties may become appreciable.  These energies
thus provide a test of the ability of the Skyrme and relativistic mean
field theories to find neutron matter properties by extrapolating data
on nuclear binding energies, sizes, vibrations, etc.  As shown in
Fig.1, the neutron matter energies predicted by the various
functionals are widely different, and not in agreement with the
results of many-body calculations at $\rho < \rho_0$.  It thus appears
likely that the simple forms of effective interactions or Lagrangians
used in the present mean field theories are inadequate to predict the
properties of neutron star matter by extrapolating the observed
nuclear properties.  Nevertheless, effective mean-field Lagrangians
have been widely used in neutron star studies due to their simplicity
\cite{glenb,Lattimer}.

\section{Many-Body Theory of Nucleon Matter}

Many properties of nuclei and nuclear matter can be understood from 
the Hamiltonian:
\begin{equation}
H = -\sum_i \frac{1}{2m} \nabla_i^2 + \sum_{i<j} v_{ij} + 
    \sum_{i<j<k} V_{ijk} + \cdots ,
\label{eq:nrh}
\end{equation}
which includes the kinetic energy, two-nucleon interactions denoted 
by lower case $v$ and three-nucleon interactions by capital $V$. 
The ellipsis denote neglected four and higher body interactions.
In this section we review the present status of this approach and its 
limitations. 

\subsection{Models of Two Nucleon Interaction}

Our understanding of QCD has not yet progressed enough to predict the 
two-nucleon interaction $v_{ij}$ {\em ab initio}.  The long range part
of $v_{ij}$ is known to be mediated by pions, the lightest of all the
mesons, and it is denoted by the one pion exchange potential (OPEP)
given by:
\beqa
v^{\pi}_{ij} &=& \frac{f_{\pi NN}^2}{4 \pi}\ \frac{m_{\pi}}{3}\ X_{ij} 
\ \boldtau_i \cdot \boldtau_j \ ,       \\
\label{eq:opep}
X_{ij} &=& Y_{\pi}(r_{ij})\ \boldsigma_i \cdot \boldsigma_j \ + \ 
T_{\pi} (r_{ij})\ S_{ij} \ .
\label{eq:xopep}
\eeqa
The pion-nucleon coupling constant $f^2_{\pi NN}/4 \pi = 0.075 \pm 0.002$ 
\cite{fpinn}, and the radial functions associated with the spin-spin and 
the tensor parts are:
\beq
Y_{\pi}(r) = \frac{e^{-x}}{x}\ \xi_Y(r) \ , \ \ \ 
T_{\pi}(r) = \left( 1 +  \frac{3}{x}  +  \frac{3}{x^2} \right) 
Y_{\pi}(r) \ \xi_T(r) \ ,
\label{eq:tpi}
\eeq
where $x = m_{\pi}r $.  The tensor operator,  
$ S_{ij} = 3\ \boldsigma_i \cdot \hat{\bf r}_{ij}
\boldsigma_j \cdot \hat{\bf r}_{ij}\ - \oss \ $, 
and since $T_{\pi}(r)\gg Y_{\pi}(r)$ in the important $x\la 1$ region, 
the OPEP is dominated by its tensor part.  
The complete $v_{ij}$ is expressed as $ v^{\pi}_{ij} + v^R_{ij} $, 
where $v^R_{ij}$ contains all the other, heavy meson, multiple meson 
and quark exchange parts.  It has to be obtained along with the short 
range cutoffs $\xi_Y(r)$ and $\xi_T(r)$ in the OPEP by fitting NN scattering 
data.  In boson-exchange models the $v^R$ is 
approximated by a sum of attractive scalar meson, and repulsive vector 
meson exchange potentials, while other models   
use attractive two-pion exchange and repulsive core potentials. 
The OPEP contains a 
$\delta-$function term omitted from Eq.(\ref{eq:opep}).  Due to 
the finite size of nucleons this term acquires a finite range, and is 
difficult to separate from $v^R$.

In the early 1990's the Nijmegen group \cite{SKR93} carefully examined
all the data on elastic NN scattering, at energies below the pion
production threshold of $\sim $ 350 MeV, published between 1955 and
1992.  They extracted 1787 proton-proton and 2514 proton-neutron
``reliable'' data, and showed that these could determine all NN
scattering phase shifts and mixing parameters quite accurately, thus
claiming that the experimental information on elastic NN scattering is
now complete.  Additional measurements are being carried out at
several laboratories including the Indiana University Cyclotron
Facility, and the CELCIUS facility in Uppsala, Sweden, to test the
accuracy of the claim and improve on the quality of the data base.
Nevertheless the Nijmegen analysis has been a major step.

NN interaction models which fit the  
Nijmegen data base with a ${\chi}^2/N_{data} \sim 1$ are called ``modern''.
These include the Nijmegen models \cite{SKT94} called Nijmegen I, II and
Reid 93, the Argonne $v_{18}$ \cite{WSS95} (A18) and CD-Bonn \cite{MSS96}.
In order to fit both the proton-proton and neutron-proton scattering data
simultaneously and accurately, these models include a detailed description
of the electromagnetic interactions and terms that violate the isospin
symmetry of the strong interaction via the differences in the masses of the
charged and neutral pions, etc.  

These five models use different parameterizations of the $v^R_{ij}$, and the
Nijmegen-I and CD-Bonn also include nonlocalities suggested by
boson-exchange representations.  Thus, like the older models, they make
different predictions for the many-body systems.  However, the differences in
their predictions are much smaller than those between older models, 
and can be partly understood.  

The interaction in the spin-isospin $T,S = 0,1$ has the largest model
dependence, which gets carried over to the estimates of the energy of 
symmetric nuclear matter (sec.~3.4).
Fortunately the deuteron structure provides significant
information on this interaction.  Fig.2 shows the deuteron wave
functions obtained with the five modern potentials \cite{ppcap}.  The
three potentials, Reid 93, Nijmegen II and A18, which are local in
each NN partial wave, give essentially the same deuteron wave
function.  We expect that they will give rather similar matter
properties.  The Nijmegen I and CD-Bonn potentials have momentum
dependent terms associated with heavy meson exchange.  These two
potentials give larger $^3S_1$ wave functions at $r < 0.8$ fm, because
they have softer repulsive cores than the local models; however, this
effect is not very large as can be seen from Fig.2.  At $r > 0.8$ fm
only the CD-Bonn predictions differ from the rest.  This is because
CD-Bonn has a strongly nonlocal OPEP as suggested by pseudoscalar
pion-nucleon coupling, which suppresses the $^3D_1$ wave function. 

These results indicate that the main difference between the preset
models of $v_{ij}$ is from the assumed nonlocality in OPEP.  However,
relativistic field theories permit use of OPEP with different
nonlocalities \cite{jununi} related with the Dyson
transformation.  The three-nucleon interaction, $V_{ijk}$, depends
upon the choice of OPEP \cite{fricoo}, and the final results obtained
after including it should be independent of the choice.  Therefore, if
relativistic field theories can be used to describe pion exchange
forces, one can use either the local OPEP in A18, or the nonlocal one
in CD-Bonn.  In this case it may be better to use the local
representation because it is simpler, and 
more accurate many-body calculations can be carried
out with it.

Realistic models of nuclear forces, the modern as well as the older, 
predict the existence of a dense toroidal inner core in
the deuteron \cite{toroid}.  The density distribution, $\rho({\bf
r}^{\prime})$, of the deuteron in spin projection $M = 0$ state, in
the center of mass frame is shown in Fig.3.  Here ${\bf
r}^{\prime} = \pm {\bf r}/2 $ are the nucleon coordinates in the
center of mass frame, and ${\bf r}$ is the internucleon distance.
This density distribution is symmetric under rotations about the
$Z^{\prime}$-axis, and the top part shows its cross section in the
$X^{\prime}-Z^{\prime}$ plane as predicted by the A18 model.  The
bottom part shows the toroidal shape of the equi-density surface for
half maximum density.  The density peaks on a ring of diameter of $\sim$ 1.0 fm
inside this torus having thickness of $\sim$ 0.8 fm.  The shape is
produced by constructive (destructive) interference between the S- and
the D-wave functions shown in Fig.2 along the $X^{\prime}$
($Z^{\prime}$) axis of the deuteron in the $M=0$ state.

The dominant, static part of the $v_{ij}$ in the deuteron, obtained by
omitting the terms dependent on the angular momentum, is anisotropic
due to the tensor part of the OPEP.  It strongly depends upon the
angle between the unit vector ${\bf r}_{ij}$ and the spin directions.
The expectation value of the interaction in the $S=1,\ M_S = 0$ state,
$\frac{1}{\sqrt{2}} |\uparrow \downarrow + \downarrow \uparrow \rangle $, 
is shown in Fig.4 as a function of $r_{ij}$ for $\theta=0$ and
$\theta={\pi}/2$.  The interaction is attractive for $\theta={\pi}/2$
and repulsive for $\theta=0$, like that between two magnetic dipoles.
The OPEP is also shown in Fig.4 by dashed lines.  The NN interaction
in all states except those with $T,S=1,0$ is dominated by the OPEP at
$r_{ij} > 1$ fm.  The OPEP is weakest in the $T,S=1,0$ states.  At
small $r_{ij}$ the repulsive core of $v^R_{ij}$ dominates in all
states.

Most of the $M=0$ deuteron wave function has also 
$M_S=0$.  Thus it is possible to understand the density distribution
of the deuteron from the potential shown in Fig.4.  The two peaks in
the density shown in Fig.3 correspond to the two-nucleons in the
deuteron being $\sim$ 1 fm apart at $\theta \sim \pi/2 $ where the
potential has its minimum value of $\sim -200 $ MeV in the A18 model.
Other models also have a deep minimum at this position.  The smallness
of the density along the $Z$-axis is due to the repulsive potential at
$\theta = 0$. Even though these features existed in the
potential models of the sixties their experimental confirmation came
via a series of measurements of the electromagnetic form factors of
the deuteron up to momentum transfers of $\sim$ 8 fm$^{-1}$ conducted
since the mid eighties at SLAC and Bates.
These and the more recent, high precision measurements
\cite{df1,df2,df3} carried out at the Jefferson Lab are in good
agreement with the predictions of the A18 model, and verify the
predicted deuteron structure beyond $r \sim 0.7$, or equivalently 
$r^{\prime} \geq 0.35 $ fm.

The $T,S = 1,0$ two-proton distribution functions, believed to be
similar to the two-neutron distribution functions due to isospin
symmetry, are predicted \cite{toroid} to have a dip at $r \sim 0 $ and
a peak at $r \sim 1$ fm, due to the repulsive core and the minima of
$v_{NN}$ respectively.  The experimental information on these
spherically symmetric distribution functions is less direct.  It comes
from sums of longitudinal response functions of light
nuclei \cite{csrmp}. The sums observed in $^3$He and $^4$He are in
fair agreement with theory, and show evidence of the predicted
structure, however, the relativistic and other corrections to the
observed sums are significant.

The observed deuteron form factors and, to a lesser extent, the sums
of longitudinal response of light nuclei indicate that the modern
two-nucleon potentials and the wave functions they predict have
validity at internucleon distances larger than $\sim$ 0.7 fm.  This
may appear surprising because the rms charge radius of the proton is
known to be 0.8 fm.  However, the nucleons seem to have a small and
dense core.  The charge form factor of the proton, as well as the
magnetic form factors of the proton and neutron are well approximated
by the dipole $(1+q^2/q_0^2)^{-2}$ with $q_0 = 840 $ MeV/c.  Inverting
this form factor gives the proton charge density as $\rho^p_{ch}(r) =
3.3\ e^{-r/0.23fm} $ fm$^{-3}$.  The charge densities of two protons,
one fm apart are shown in Fig.5.  It should be noted that the charge
densities shown can have corrections due to the neglect of
relativistic effects in inverting the form factor at $r\la 1/m \sim
0.21 $ fm, and that recent measurements of the proton form factor
\cite{protff} show deviations from the dipole form at momenta $\ga$ 4 
fm$^{-1}$. They suggest that proton charge density flattens out at $r < 0.3$
fm. Nevertheless Fig. 5 indicates that nucleons a fm apart can retain 
their identities.  The NN interaction includes the change in the 
energy due to their overlap, and it has minima near $r \sim 1$ fm.  In
absence of the quantum kinetic energy term $( - \nabla^2/2m)$ in the
Hamiltonian (Eq.\ref{eq:nrh}) the deuteron will shrink to a ring of
radius $\sim$ 0.5 fm, and the equilibrium density of nuclear matter
will be $\sim$ 1 fm$^{-3}$.  The density of matter in most neutron 
stars is less than that. 

\subsection{Models of Three Nucleon Interaction}

All realistic models of $v_{ij}$, the modern and the older, underbind the 
triton and other light nuclei and predict too high equilibrium density for 
symmetric nuclear matter.  In both cases the deviation from experiment is 
not too large, particularly when compared with the expectation values 
of $v_{ij}$.  For example, the expectation value 
$\langle v_{i < j}  \rangle $ in $^3$H 
is about $\sim - 50 $ MeV, while the underbinding is by $<$ 1 MeV.  
It is likely that these differences are due to three-nucleon interactions 
expected and predicted since the fifties \cite{fumi}.

High precision modeling of the two-nucleon interaction is possible because 
scattering cross sections can be easily and exactly calculated from the 
assumed $v_{ij}$, and a complete set of $\sim$ 4000 cross sections has been 
measured.  Such an approach is not practical at present for the three-nucleon 
interaction $V_{ijk}$.  In principle deuteron-nucleon scattering and 
reactions can be used to study $V_{ijk}$ in the isospin $T = 1/2$ state.  
However, this scattering is dominated by the two-nucleon interaction 
\cite{glodns}, and very high precision data is necessary to extract 
the effects of $V_{ijk}$, as is being attempted in new experiments at 
the Indiana University cyclotron, focused on spin observables. 

The large $ \langle v_{i<j} \rangle $ is cancelled to a large extent 
by the kinetic energy in nuclear binding energies.  Thus $ \langle 
V_{i<j<k} \rangle $ in nuclei is expected to be of the order of 10 $\%$ of their 
binding energy.  Hence we can construct realistic models 
of $V_{ijk}$ by fitting binding energies of light nuclei, which 
can now be calculated with 
an accuracy of the order of 1 $\%$ using Greens Function Monte Carlo (GFMC)
methods \cite{pud7,next8} and the estimated equilibrium properties of nuclear 
matter.  As will be discussed in sec.~3.7, it is not yet possible 
to calculate the EOS of symmetric nuclear matter with comparable accuracy. 
The models depend upon the 
$v_{ij}$ used in the Hamiltonian.  This is inevitable, 
because unitary transformations make correlated changes in $v_{ij}$ 
and $V_{ijk}$ \cite{fricoo}.  Only the combinations of $v_{ij}$ and 
$V_{ijk}$ in the Hamiltonian (\ref{eq:nrh}) are meaningful.  

The information contained in nuclear binding energies and equilibrium 
properties of nuclear matter is limited.   Therefore the realistic 
models of $V_{ijk}$ rely on theory to a
much larger extent than the models of $v_{ij}$, and contain very few 
parameters.  The Urbana models of $V_{ijk}$ contain two isoscalar terms:
\begin{equation}
V_{ijk}=V^{2\pi}_{ijk} + V^{R}_{ijk} \ .
\label{eq:uvijk}
\end{equation}
The first term represents the Fujita-Miyazawa \cite{fumi} 
two-pion exchange interaction:
\beq
V^{2\pi}_{ijk}=\sum_{cyc} A_{2\pi} \left( \left\{\ott,
\boldtau_i \cdot \boldtau_k \right\}
\left\{ X_{ij},X_{ik} \right\}
+ \frac{1}{4} [\ott, \boldtau_i \cdot \boldtau_k]
[X_{ij},X_{ik}] \right),  
\label{eq:fumi}
\eeq
with strength denoted by 
$A_{2\pi}$. The functions $T_{\pi}(r_{ij})$ and $Y_{\pi}(r_{ij})$
in $X_{ij}$, Eq.(\ref{eq:xopep}), are taken from the A18 
model of $v_{ij}$. This interaction is due to the pion exchanged by nucleons 
$j$ and $k$ being scattered by the nucleon $i$ via the $\Delta$ resonance in 
$\pi$-N scattering.  In classical terms it is due to the polarization of the 
quark spins in nucleon $i$, due to the pion field of $j$ ($k$), 
interacting with $k$ ($j$); and it is similar 
to the three-body earth-moon-satellite gravitational interaction due to the 
polarization of the ocean water on earth by the moon's gravity. 

The $V^R_{ijk}$ is purely phenomenological, and has the form:
\begin{equation}
V^{R}_{ijk}=U_0\sum_{cyc}T_\pi^2(r_{ij})T_\pi^2(r_{ik}).
\end{equation}
It was meant to represent the modification of the two-pion 
exchange part of $v_{ij}$  
by other particles in matter, however, $\sim$ 40 $\%$ of it is due to 
relativistic effects discussed in the next subsection. 
The parameters of the
present model U-IX, $A_{2\pi}=-0.0293$ MeV and $U_0=0.0048$ MeV, have
been determined from exact GFMC calculations of $^3$H and approximate
variational calculations of the equilibrium density of 
nuclear matter with the A18 NN interaction \cite{Akma97}. 

The results of essentially exact GFMC calculations \cite{pud7,next8}
with the A18 and U-IX interactions are shown in Table-\ref{tb:qmc}.
We note that the better known pion exchange parts of these
interactions give the largest contributions, but the contributions of
the phenomenological parts, $v^R_{ij}$ and $V^R_{ijk}$, are
significant.  The column $\Delta E_{expt}$ gives the difference
between the experimental energies and the calculated, while $\Delta
E_{VMC}$ is that between the variational Monte Carlo (VMC) upper
bounds and the exact GFMC energies.

\begin{table}
\caption{Results of Quantum Monte Carlo Calculations in MeV}
\vspace{0.2cm}
\begin{tabular}{rlrrrrrrr}
\hline
$^A$Z & $(J^{\pi};T)$   &  $v^{\pi}_{ij}$  &  $V^{2{\pi}}_{ijk}$  &  $v^R_{ij}$  &  $V^R_{ijk}$  &  $
E_{GFMC}$  &  ${\Delta}E_{expt.}$  &  ${\Delta}E_{VMC}$\\
\hline
$^2$H & $(1^+;0)$ & -21.3 & 0 & -0.8 & 0 & -2.22 & 0 & 0 \\
$^3$H & $({\frac{1}{2}}^+;\frac{1}{2})$ & -43.8 & -2.2 & -14.6 & 1.0 & -8.47 & -0.01(1) & 0.15 \\
$^4$He & $(0^+;0)$ & -99.4 & -11.7 & -36.0 & 5.3 & -28.30 & 0.00(2) & 0.52 \\
$^6$He & $(0^+;1)$ & -109 & -13.6 & -56 & 6.4 & -27.64 & -1.63(14) & 2.8 \\
$^6$Li & $(1^+;0)$ & -129 & -13.5 & -50 & 6.3 & -31.25 & -0.74(11) & 3.2 \\
$^7$He & $({\frac{3}{2}}^-;\frac{3}{2})$ & -110 & -14.1 & -61 & 6.7 & -25.2 & -3.7(2) & 4.7 \\
$^7$Li & $({\frac{3}{2}}^-;\frac{1}{2})$ & -153 & -17.1 & -68 & 8.2 & -37.4 & -1.8(3) & 4.7 \\
$^8$He & $(0^+;2)$ & -121 & -15.8 & -74 & 7.5 & -25.8 & -5.6(6) & 6.1 \\
$^8$Li & $(2^+;1)$ & -157 & -22.2 & -104 & 11.0 & -38.3 & -3.0(6) & 8.6 \\
$^8$Be & $(0^+;0)$ & -224 & -28.1 & -72 & 13.3 & -54.7 & -1.8(6) & 6.6 \\
\hline
\end{tabular}
\label{tb:qmc}
\end{table}

Table-\ref{tb:qmc} shows that the U-IX interaction underbinds $A=8$
nuclei, and since $^8$He is more underbound than $^8$Be, it
misrepresents the isospin dependence of $V_{ijk}$.  The new Illinois
models of $V_{ijk}$ \cite{nextV} resolve this problem by including the
leading three-pion exchange term, $V^{3\pi}_{ijk}$, that is attractive
in triplets having isospin $T=3/2$, but has little effect on the
$T=1/2$ triplets in $^3$H and $^4$He.  A much improved fit, with
errors $< 2 \%$, to the observed energies is obtained 
as shown in Fig.6.  The three parameters
in Pieper's model IL-2R are the strengths of the $V^{2\pi}_{ijk}$,
$V^R_{ijk}$ and $V^{3\pi}_{ijk}$.  Calculations of NM properties with
this more accurate $V_{ijk}$ are in progress; the preliminary results
are similar to those with U-IX since the $V^{3 \pi}$ is much weaker
than the $V^{2 \pi}$ and $V^R$.  For example, the expectation values
of $V^{2\pi}_{ijk}$, $V^R_{ijk}$ and $V^{3\pi}_{ijk}$ of IL-2R model,
in $^8$Be ($^8$He) are respectively $\sim$ $-$38 ($-$27), +19 (+14)
and $-$2 ($-$5) MeV.  In the following sections we will review the
properties of neutron star matter calculated with A18 and U-IX
interactions.

In principle there can be four-nucleon interactions (FNI) neglected in
the Hamiltonian (Eq.\ref{eq:nrh}).  It seems that they are very weak
in nuclei.  All the models of $V_{ijk}$ studied so far reproduce the
energy of $^4$He with an error $< 0.5 \% $, after fitting the observed
energy of $^3$H.  Since this error is close to the accuracy of the
$^4$He calculation, there is no indication of FNI in that nucleus.
The IL-2R model also gives the experimental energies of $A=8$ nuclei
within $\sim 1 $ MeV.  In $^8$Be, for example, the expectation values
of $v_{ij}$ and $V_{ijk}$ are respectively $-$308 and $-$21 MeV
respectively.  By comparison with experiment we estimate that the
possible contribution of FNI in this nucleus is $<$ 1 MeV.  The
$V_{ijk}$ presumably has additional smaller terms neglected in IL-2R,
but it is difficult to determine their strengths from the nuclear
spectra that can be calculated accurately from bare forces at present.

\subsection{Relativistic Boost Interaction}

In all models, the NN scattering data is reduced to the center of mass
frame and fitted using phase shifts calculated from the NN
interaction, $v_{ij}$, in that frame.  The $v_{ij}$ obtained by this
procedure describes the interaction between nucleons having total
momentum ${\bf P}_{ij} = {\bf p}_i + {\bf p}_j = 0 $.  In general, the
interaction between particles depends on their momenta ${\bf p}_i$ and
${\bf p}_j$.  For example the 
electromagnetic interaction \cite {BS57} between two
particles of mass {\em m} and charge {\em Q}, contains a term with
the factor $ -{\bf p}_i \cdot {\bf p}_j /2m^2 = p^2_{ij}/2m^2 -
P^2_{ij}/8m^2 $, where ${\bf p}_{ij} = ({\bf p}_i - {\bf p}_j)/2$ is
the relative momentum.  The terms containing ${\bf p}_{ij}$ are included
in the momentum-dependent parts of $v_{ij}$, while the boost interaction 
${\delta}v({\bf P}_{ij})$ contains parts dependent on the total
${\bf P}_{ij}$.  Even though
contributions of the boost interaction to the binding energy of SNM
and $^3$H were estimated by Coester and coworkers in the seventies and
eighties \cite{CPS74,GLC86}, they were first included in studies of
dense matter rather recently \cite{apr98}.  Walecka's relativistic
mean field theory naturally contains the boost interactions \cite{FPF95}.

Following the work of Krajcik and Foldy, Friar \cite{Fri75}
obtained the following equation relating the boost interaction of order
$P^2$ to the interaction in the center of mass frame:
\begin{equation}
{\delta}v({\bf P}) = -\frac{P^2}{8m^2} v
+\frac{1}{8m^2}[{\bf P \cdot r \; P \cdot {\nabla}},v]
+\frac{1}{8m^2}[({\bf \sigma}_i - {\bf \sigma}_j) \times
{\bf P \cdot \nabla}, v] .
\label{eq:friar}
\end{equation}
The general validity of this equation in relativistic mechanics and field
theory was recently demonstrated \cite{FPF95}.
Including boost interaction, the nonrelativistic
Hamiltonian assumes the form:
\begin{equation}
H^\ast = \sum \frac{p^2_i}{2m} + \sum (v_{ij} + {\delta}v({\bf P}_{ij}))
+ \sum V^*_{ijk} + \cdots ,
\label{eq:nrhwb}
\end{equation}
where the ellipsis denotes the three-body boost, and four and higher body
interactions. This $H^\ast$ contains all terms quadratic in the particle
velocities, and is therefore suitable for complete studies in the
nonrelativistic limit.

Studies of light nuclei using the VMC method \cite{CPS93,For98} find
that the contribution of the two-body boost interaction to the energy
is repulsive, with a magnitude which is $\sim$ 37\% of that 
of $V^R_{ijk}$ in the UIX model.
The boost interaction thus accounts for a significant
part of the $V^R_{ijk}$ in Hamiltonians which fit nuclear energies
neglecting ${\delta}v$.  The $V^*_{ijk}$ in Eq.(\ref{eq:nrhwb}) has
a $V^R$ of appropriately smaller strength than that in the $V_{ijk}$
in Hamiltonian $H$ given by Eq.(\ref{eq:nrh}).

We should expect additional relativistic corrections to the
Hamiltonian (\ref{eq:nrhwb}).  However, when nonrelativistic
potentials are fit to the experimental data, relativistic effects
present in the data are automatically buried in these potentials.  In
order to study the magnitude of a chosen relativistic correction, such
as that due to the approximation of the kinetic energy or the boost
interaction, or the nonlocalities of OPEP, it is necessary to refit the
same data set, fitted to obtain the nonrelativistic Hamiltonian, and
then study the differences.  Such comparisons \cite{For98} indicate
that the relativistic corrections associated with kinetic energies and
nonlocalities of OPEP are small, whereas the boost
corrections are significant.  This is not surprising since the boost
interaction was totally omitted from the conventional nonrelativistic
nuclear Hamiltonian (Eq.\ref{eq:nrh}).  

\subsection{Brueckner Calculations of Nucleon Matter}

Calculating the properties of matter from the interaction $v_{ij}$
between pairs of its constituents is a well known problem in many-body
theory.  It is particularly challenging for NM due to the
strong spin-isospin dependence of $v_{ij}$.  In the method developed
by Brueckner, Bethe and Goldstone the perturbation expansion of the
energy of NM is cast into a series ordered according to
the number of independent hole lines (HL).  This method has been used
extensively since the sixties \cite{bethe71} to study symmetric
nuclear matter (SNM) and nuclei.  It was also used then to predict
properties of pure neutron matter (PNM) \cite{siepan71} soon after 
the discovery of pulsars.  The
convergence of the expansion depends upon the choice of the
single-particle energies in the assumed unperturbed Hamiltonian.  For
the hole states with momenta less than $k_F$, they are chosen self
consistently as suggested by Brueckner and Gammel \cite{brugam58} via
the Brueckner-Hartree-Fock (BHF) procedure.  The older calculations by
Day \cite{dayw85} used kinetic energies for particle states.  Since
this leads to a discontinuity in the unperturbed single-particle
energies at $k_F$, Day's choice is called ``discontinuous''.  In 1976
the Li\`{e}ge group \cite{liege76} advocated the ``continuous'' choice
by extending the definition of BHF single-particle energies to particle
states.  If all the higher order terms of the HL expansion are
computed the final results should be independent of the choice.

Detailed calculations of SNM have been carried out with the older Argonne
$v_{14}$ interaction by the Catania group
\cite{catprl98} using both choices.  The results of the lowest order
2-HL calculation depend significantly on the choice, however, those
including 2+3-HL terms are almost the same for the two choices.  They
find that the 2-HL results with the continuous (discontinuous) choice
are $\sim$ 15 \% below (30 \% above) those of the 2+3.  Day's
calculations \cite{dayw85}, with the same interaction, include
additional 4-HL terms, and give energies below the 2+3 Catania
results, closer to the 2-HL continuous.  For example, the energies of
SNM at $\rho = 0.28$ fm$^{-3}$ are $-$11.3, $-$16.1, $-$18.3 and $-
17.8 \pm 1.3$ for the Catania 2-HL discontinuous, 2+3-HL, 2-HL
continuous, and Day's calculations respectively, while the variational
{\em upper bound} at this density is $-16.2 \pm 0.4$ \cite{dayw85} with
the methods described in the next subsection.  
The size of the error due to truncation
of the expansion is estimated in Day's and variational calculations.

A significant advantage of Brueckner's method is that it can be easily
applied to local as well as nonlocal modern interactions.  The present
calculations, called lowest-order BHF (LOBHF), include only the 2-HL
terms and use the continuous choice \cite{Eng97}.  Their results for
PNM and SNM are shown in Fig.7; those for matter with intermediate
values of proton fraction can be estimated from these by interpolating
with $\beta^2$ as metioned in sec.~2.  These results provide an
estimate of the uncertainty in the predicted matter energy due to that
in the NN interaction.

The LOBHF energies for neutron matter are essentially model
independent up to $\sim 0.3$ fm$^{-3}$ (Fig.7); at higher densities
they deviate partly for the following reason.  All the models fit the
NN scattering data up to 350 MeV lab energy, {\em i.e.} up to maximum
relative momenta $k = 2.05$ fm$^{-1}$.  The maximum value of the
relative momentum of two hole states in matter is $k_F$, and the
density of PNM (SNM) at $k_F = 2.05$ is 0.29 (0.58) fm$^{-3}$.  In PNM
at $\rho > 0.29$ the interactions are being used at relative momenta
larger than in the fitted data.  Baldo {\em et al} \cite{baldo98} have
shown that the $^3P_2$ phase shifts predicted by the five modern
potentials vary from 8 (A18) to 19 (Nijmegen II) degrees at $k = 3$
fm$^{-1}$.  The average value, $k_{rms} = \sqrt{3/10}\ k_F$, is
smaller than $k_F$, and exceeds 2.05 fm$^{-1}$ at much larger
densities of 1.77 (3.54) fm$^{-3}$ of PNM (SNM).  Presumably, this
helps to keep down the model dependence.

The LOBHF energies of SNM have a larger model dependence starting at lower 
densities (Fig.7).  Here 
the main cause seems to be the assumed nonlocality of $v_{ij}$ discussed 
in sec.~3.1.  The local 
interactions, Nijmegen II, Ried 93 and A18 give similar results, while the 
most nonlocal CD-Bonn gives the lowest energies. 
The predicted values of equilibrium $\rho_0$ and $E_0$ of SNM
are respectively 0.31, 0.27, 0.28, 0.27 and 0.37 fm$^{-3}$
and $-20.3,\ -17.6,\ -18.7,\ -18.1$ and $-22.9$ MeV
with Nijmegen I, II, Reid93, A18 and CD Bonn interactions;
while the empirical values are 0.16 fm$^{-3}$ and $-16$ MeV.
Obviously, the empirical properties of SNM can not be obtained by
approximating the nuclear interaction energy by $ \sum v_{i<j}$.
TNI, added to the Hamiltonian to obtain the observed 
properties,  naturally depend upon the choice
of $v_{ij}$.  For example, those to be added to CD Bonn have to have
stronger repulsive parts, which dominate at large $\rho$. 
The combinations of $v_{ij}+V_{ijk}$, constrained with
experimental data, will have smaller model dependence than seen in Fig.7.

Baldo, Bombaci and Burgio \cite{baldo97} have carried out LOBHF
calculations with the older Paris and Argonne
interactions including Urbana TNI.  Like the modern
models, these older models give too large $\rho_0$ ($\sim
0.26$~fm$^{-3}$) and $E_0$ $(\sim -18$~MeV) in LOBHF without TNI.  By
averaging over the position of the third nucleon the TNI is expressed
as a density dependent NN interaction to be added to the $v_{ij}$.
The parameters $A_{2\pi}$ and $U_0$ were chosen to get closer to the
empirical values of $\rho_0$ and $E_0$; their values of $-$0.0329 and
0.00361 MeV are not too far from those of U-IX ($-$0.0293 and 0.0048).
With the Paris + Urbana model they obtain $\rho_0=0.176$~fm$^{-3}$, $E_0 =
-16.0$~MeV, $K = 281$~MeV, and $E_{sym} = 33$~MeV.  These values, as well 
as those obtained using the Argonne interaction instead of Paris, compare
rather well with the empirical values given in sec.~2.  

Relativistic effects are included in the LOBHF calculations via
Dirac-Brueckner approximation suggested by Celenza and Shakin
\cite{celsha}.  The calculations include
contributions of the boost interactions as well as TNI \cite{bwbs},
and many nucleon interactions generated via the 
$Z$-graphs containing anti-nucleon
lines.  However, they do not contain contributions of the
Fujita-Miyazawa and other TNI due to internal structure of nucleons.
Results have been reported by Brockmann and
Machleidt \cite{bromac} for the older Bonn meson exchange NN
interaction models; those with the Bonn-A model come close to 
reproducing the empirical properties of SNM at $\rho_0$.

\subsection{Variational Calculations of Nucleon Matter}

Variational calculations of NM with realistic interactions
have been carried out since 1970 \cite{vrpsol74}.  The present
calculations \cite{Akma97,apr98} use variational wave-functions, $\Psi_v$,
consisting of a symmetrized product of pair correlation operators,
$F_{ij}$, operating on the Fermi gas wave-function. In PNM, the
$F_{ij}$ include four terms generating spatial, $\boldsigma_i \cdot
\boldsigma_j $, tensor and spin orbit correlations.  The SNM $F_{ij}$
have eight terms; the additional four have $\boldtau_i \cdot
\boldtau_j$ factors.

This wave-function is clearly too simple to accurately describe the
ground state of nuclear matter. Monte Carlo studies of few-body nuclei
use additional three-body
correlations induced by both $v_{ij}$ and $V_{ijk}$, in the
variational wave-function; they reduce the energy of $^{16}$O by 
$\sim$ 1 MeV/nucleon. The results shown in
Table.\ref{tb:qmc} indicate that the VMC energies of $A=8$ nuclei,
obtained after including three-body correlation operators, are above
the exact GFMC values by $\sim$ 1 MeV/nucleon.  From these we estimate
that the present $\Psi_v$ may underbind SNM by a few MeV. In contrast,
the three-body correlations have a smaller effect on the energy
of pure neutron drops \cite{pscppr}.  The variational energy of a drop
with eight neutrons, calculated with the simple $\Psi_v$, is greater
than the exact value by $\sim$ 0.5 MeV/nucleon. Thus the variational
energies are relatively more accurate for PNM than for SNM. This is as
expected, since SNM has much stronger tensor correlations.
Despite the aforementioned
shortcomings, the simple $\Psi_v$ having only pair correlation
operators describes the gross features of the nuclear wave-function
rather well.  For example, the spin-isospin dependent two-nucleon
distribution functions calculated in this approximation are close to
the exact distribution functions \cite{pud7}.

The correlation operators $F_{ij}$ are determined from Euler-Lagrange
equations that minimize the two-body cluster contribution
of an interaction $\bar{v}_{ij}-\lambda_{ij}$.  The interaction
$\bar{v}_{ij}$ is related to the $v_{ij}$ via a parameter $\alpha$
meant to simulate the quenching of the spin-isospin interaction
between particles i and j, 
via their interaction with other particles in matter. 
The operator $\lambda_{ij}$ simulates screening effects in
matter; it is determined from the ranges $d_t$ of tensor correlations,
and $d_c$ of all the other correlations.  The  
$\Psi_v$ thus depend on three variational parameters:
$\alpha$, d$_c$ and d$_t$, determined by minimizing the energy. Two
additional parameters are used in Ref.\cite{WFF} to further lower the
variational upper bound by small amounts.

The energy expectation value is evaluated using cluster expansion.
The one-body term is just the Fermi gas kinetic
energy, and the large two-body (2B) term, analogous to the interaction
energy in LOBHF, is calculated exactly.  The most important of the
many-body (MB) cluster contributions are summed using
Fermi-hypernetted chain (FHNC) and single operator chain
equations \cite{Akma97,apr98}, and constrains are imposed to
satisfy the fundamental identities of pair distribution functions.
The kinetic energy can be calculated using different expressions
related by integration by parts. If all MB contributions are
calculated, these expressions yield the same result. However, they
yield different results when only selected parts of the MB clusters
are summed using chain equations.  Studies of atomic helium liquids
with FHNC summation methods find the exact result to be
between the energies obtained using the Jackson-Feenberg (JF) and
Pandharipande-Bethe (PB) expressions. The average of these two is used
as the result with half the difference as an estimate of the error.

More general pair correlations can be calculated by separately
minimizing the two-body cluster contribution to each partial wave,
specified by $l,S,J$ and the relative momentum $k$ \cite{pb}.  These
correlations $f(l,S,J,k)$ depend on all the quantum numbers, and yield
a lower 2B energy than the $F_{ij}$ operator with the same $\alpha$,
$d_c$ and $d_t$.  The MB contributions cannot be easily calculated
with the general $f(l,S,J,k)$, however.  The differences between
optimum $f(l,S,J,k)$ and $F_{ij}$ can be included via the second order
two-particle, two-hole contribution, $\Delta E_2$, in correlated basis
perturbation theory \cite{ffpop,ff}.  In recent calculations
\cite{Akma97,apr98} the $\Delta E_2$ is approximated by the difference
$\delta E_{2B}$ between the 2B cluster energies calculated using
$f(l,S,J,k)$ and $F_{ij}$. However, the values of $\alpha, d_c, d_t$
are determined by minimizing the energy calculated from the $F_{ij}$.

The variational \cite{apr98} and LOBHF energies obtained with the A18
interaction are compared in Fig.8.  At densities below 0.6 fm$^{-3}$
there is fairly close agreement between them, however, we expect the
true results to be a few MeV below the variational upper bound.  At
higher densities the SNM LOBHF energy is significantly below the
variational bound.  The convergence of the HL expansion is expected to
deteriorate at higher densities, and it may be the cause of the large
difference.

\subsection{Neutral Pion Condensation}

Variational calculations with the Hamiltonian (\ref{eq:nrhwb}) with 
A18+$\delta v$+UIX$^*$ interactions, indicate the occurrence of a phase
transition in both PNM and SNM; the energies of the two phases are
shown in Fig.9.  The tensor correlations have a longer range in the
higher density phase (HDP) than in the low density phase (LDP).
Detailed studies of the pion fields in the two phases \cite{Akma97}
indicate that the HDP has a large enhancement of virtual neutral pions
with momenta $\sim 1.5$ fm$^{-1}$, and therefore this transition is
believed to be due to neutral pion condensation in matter.  Although
the effect of this type of transition on the EOS is
relatively small, it can have important consequences for the cooling
and evolution of neutron stars \cite{pethick}.

Since the pioneering work of Migdal \cite{migdal} and of Sawyer and
Scallapino \cite{ss}, many investigators have used effective
interactions to study the possibility of pion condensation in SNM and
PNM.  These efforts were recently reviewed by Kunihiro {\it et al}
\cite{kun}.  Neutral pion condensation occurs when PNM (SNM) becomes
unstable towards the development of a spin (spin-isospin) density
wave as discussed in sec.~4.2.  
The variational wave functions used in the Urbana calculations
\cite{Akma97} are not adequate to describe the long range order
expected with $\pi^0$-condensation; better calculations may be
possible with the quantum Monte Carlo method described in the next
subsection.  NM with spin or spin-isospin density wave
naturally has a pion field of the same wavelength; it causes $N
\rightleftharpoons \Delta $ transitions, which help lower the energy
of this phase.  The effects of the $\Delta$-resonance are absorbed
into the interactions in the Hamiltonian (\ref{eq:nrhwb}).  In
particular, the Fujita-Miyazawa $V^{2\pi}_{ijk}$ has a large effect on
$\pi^0$ condensation.  Without it there is no condensation predicted
in SNM, while that in PNM occurs at a higher density of $\sim$ 0.5
fm$^{-3}$.

This transition was found by Wiringa, Fiks and Fabrocini 
to occur with the older Argonne
$v_{14}$ for PNM but not for SNM \cite{WFF}, while it does not occur
in either PNM or SNM with the Urbana $v_{14}$ interaction of 1981.  In
Migdal's approach \cite{migdal}, the transition of SNM to the pion
condensed phase is inhibited by a positive, short-range $\oss\ott$ NN
interaction whose strength is represented by the Landau parameter
$g^\prime$.  In the case of PNM the sum of the $\oss$ and $\oss\ott$
interactions occurs since $\ott = 1$.  The contact part of OPEP 
gives a negative $\oss\ott$ potential at small $r$, and thus favors
pion condensation.  The $\oss\ott$ part of the modern A18 NN
interaction does become negative at small $r$, however, it is
positive in the older Urbana-Argonne models which do not fit the NN
scattering data as well as the A18.

The energies of $\beta$-stable LDP and HDP phases have been calculated 
by interpolation between PNM and SNM \cite{apr98}.  The LDP phase with 
$\rho=0.204$ fm$^{-3}$ and proton fraction $x_p=0.073$ is found to be in 
equilibrium with the HDP at $\rho=0.237$ and $x_p=0.057$.  In between 
there will be mixed phase regions as discussed in sec.~5. 

\subsection{Quantum Monte Carlo Calculations}

The variational energy of the ground state of SNM, calculated with 
the A18+$\delta v$+UIX$^*$ model is $\sim -12$~MeV,
against the empirical value of $-16$~MeV.  As mentioned earlier, this
difference is believed to be mostly due to the inadequacy of the
present variational wave functions.  Accurate calculations of the
energies of many-body systems in which the interactions do not depend
upon the spins of the particles have been carried out using quantum
Monte Carlo (QMC) methods \cite{lthe}.  In these systems 
one can work with a wave
function $\Phi({\bf R})$ that depends only upon the positions of all
the particles represented by the configuration vector ${\bf R}={\bf
r}_1,{\bf r}_2,...{\bf r}_A $.  A VMC calculation is used to obtain a
good approximation $\Phi_v({\bf R})$, and in the GFMC method one
operates on $\Phi_v({\bf R})$ with the imaginary time evolution
operator $exp(-[H-E_0] \tau)$ to project out the exact ground state
$\Phi_0({\bf R})$.

The main difficulty in applying QMC methods to nuclear problems is
that nuclear forces change spins and isospins of the interacting
nucleons, and thus nuclear wave functions contain superpositions of
all possible spin-isospin states of $A$-nucleons.  
Their number, $\sim 2^A\ A!/[Z!(A-Z)!]$, increases very rapidly with $A$.  
For this reason exact QMC calculations have been carried out only for
nuclei having up to 8 nucleons \cite{next8,nextV}, and attempts to
calculate $A=9$ nuclei are in progress.  Carlson \cite{joe14n} has
also calculated the ground states of 14 neutrons in a periodic box.
Calculations of pure neutron systems are simpler because there is only
one isospin state, and all $\ott$ can be replaced by the unit
operator.

In addition, GFMC calculations of Fermion systems suffer from the
``Fermion sign problem''. The real wave functions of simple Fermi
systems have nodal surfaces because the $\Phi({\bf R})$ must equal $-
\Phi({\bf R}^{\prime})$ when the configurations ${\bf R}$ and ${\bf
R}^{\prime}$ are related by the exchange of a pair of particles. GFMC
configurations which diffuse across nodal surfaces, as the system
evolves in imaginary time, increase the variance of the calculated
quantity, making unconstrained propagation impractical for large
systems. In the fixed node method \cite{fixnd} for simple systems
this growth of variance is eliminated by restricting the
configurations to domains enclosed by the nodal surfaces of the
variational wave function.  Such calculations generally have an error
due to imperfections in that structure, but it is much smaller than
that in variational.

A similar problem comes in nuclear GFMC with the additional complexity
due to nuclear wave functions having many spin-isospin components,
each with a different nodal structure.  The growth of the variance
is tolerable for $A \leq 7$, but when $A \geq 8$
it is necessary to use constrained path methods \cite{next8} to
control the variance.  The constraint can be removed at large $\tau$
to test if it influenced the calculated energy significantly.  It
appears that calculations with $\sim$ 2\% accuracy in the binding
energy are possible in this way for systems having up to 14 neutrons.

Auxiliary field diffusion Monte Carlo (AFDMC) \cite{afdmc} seems to be
the long-sought breakthrough needed to eliminate the exponential
$(2^A)$ growth of spin states in GFMC calculations of neutron matter.
This method combines two major themes in QMC.
Auxiliary-field methods are used in the shell model Monte Carlo
calculations \cite{smmc}, and several condensed matter systems in
which the continuous spatial degrees of freedom have been eliminated, 
while diffusion Monte Carlo is another name for GFMC.

In the approach developed by Schmidt and Fantoni \cite{afdmc}, the
spatial parts, ({\em i.e.} kinetic energy and spin-independent
interactions), of the Hamiltonian are propagated as in GFMC and the
spin-dependent interactions between neutrons are replaced by
interactions of neutrons with auxiliary fields. Integrating over the
auxiliary fields reproduces the original spin-dependent interaction.
In addition, a constraint analogous to the fixed-node approximation in
GFMC is introduced, by requiring that the real part of the overlap
with a trial function remains positive.

More recently, Schmidt and Fantoni \cite{sfpriv} have
carried out calculations with a realistic Hamiltonian consisting
of Argonne $v_8^{\prime}$ NN interaction used in the GFMC calculations
\cite{pud7}, and the UIX TNI. The $v_8^{\prime}$ contains the main
parts of A18, and the difference between the two is treated
perturbatively.  Results of calculations with 38 neutrons in a
periodic box with finite size corrections have been obtained. They are
$\sim 5 \%$ below the variational energies obtained with the
methods described above.  For example, the AFDMC and variational
energies for Argonne $v_8^{\prime}$ and UIX interactions are 21.8
$\pm 0.1$ (65.5 $\pm 0.1$) and 23.2 (68.6) MeV per neutron at $\rho =
0.2 \ (0.4)$ fm$^{-3}$.

The trial functions used to constrain the present AFDMC calculations
are rather simple without any spin correlations.  In contrast it is
possible to use more accurate variational wave functions with spin
correlations to constrain the GFMC calculations.  Carlson
\cite{joe14n} has compared AFDMC and GFMC results for 14 neutrons in a
periodic box.  At $\rho = 0.15$ fm$^{-3}$ the GFMC energy (220 $\pm$
1) is about 7 \% below the AFDMC result of 236.4 $\pm$ 1.5 MeV.  From
these we conclude that the variational PNM energies given in the last
section may be $\sim$ 12 \% above the exact values for the A18 and
UIX interactions.  The error in SNM $E(\rho)$ is probably twice as
large.  As mentioned in the conclusions (sect.7), an overestimation 
of the $E(\rho)$ of neutron matter 
by 12 \% has a rather small effect on the predictions 
of neutron star properties. 

The AFDMC is more accurate than the present variational method, and 
it is also more versatile.  For example, it 
can be used to study matter with long range spin-isospin order
induced by $\pi^0$-condensation discussed in sect 4.2.

\section{Hadronic and Quark Matter}

It is likely that at high densities 
more general form of matter containing hadrons besides the nucleons,  
called hadronic matter (HM), has lower energy.  
The possibilities are that it contains  
negatively charged mesons like pions or kaons, 
or other hyperons such as $\Sigma^-$ or $\Lambda$. 
Finally, it is expected that 
at a high enough density there will be a transition 
to quark matter (QM) in which the quarks are 
not clustered into nucleons or hadrons.  

The interactions between hyperons and nucleons, and between kaons 
and nucleons are not as well known as those between nucleons, and 
the energy of quark matter is difficult to calculate realistically. 
Therefore the transition densities from NM to HM or QM 
are rather difficult to calculate reliably.  We review the 
recent estimates. 

\subsection{Kaon Condensation}

Kaon condensation in dense matter was suggested by Kaplan and Nelson
\cite{kn87}, and has been discussed in many recent publications
\cite{BLRT,Weise}. Due to the attraction between $K^-$ and nucleons
the kaon energy decreases with increasing density, and
eventually if it drops below the electron chemical potential in
NM, a Bose condensate of $K^-$ will appear.  
The key quantities of interest are the electron and kaon 
chemical potentials in NM.  The former is obtained 
from the $\beta$-equilibrium condition, $\mu_e = \mu_n - \mu_p $ 
relating electron, neutron and proton chemical potentials.
The $\mu_e$ obtained from the A18+$\delta v$+UIX$^*$ 
interactions with variational calculations is shown in Fig.10. 

In neutron matter at very low densities, the interparticle spacing is much
larger than the range of the $K^-n$ interaction, and 
the kaon interacts many times with the same nucleon before it
encounters and interacts with another nucleon. 
Therefore one can use the scattering length, $ a_{K^-n}$, 
as the ``effective'' kaon-nucleon
interaction. In this low density limit the kaon energy deviates from its 
rest mass by the Lenz potential, and is given by \cite{kaon} :
\begin{equation}
   \omega_{Lenz} = m_K + \frac{2\pi}{m_R}\, a_{K^-n}\,\rho ,
               \label{Lenz}
\end{equation}
where $m_R = m_K m_n /(m_K + m_n)$ is the kaon-neutron reduced mass. 
The scattering length extracted from data is $(-0.37 -i0.57)$ fm; 
its imaginary part is due to the open $\Lambda-\pi^-$ channel in 
vacuum.  In the density region of interest to kaon condensation 
the kaon energy is too small for the 
$K^-+n \rightarrow \Lambda + \pi^-$ reaction to occur.  Using effective 
lagrangians based on chiral perturbation theory Brown {\em et. al.} 
\cite{BLRT} estimate $a_{K^-n}$ to be $-0.41$ fm in absence of reaction channels. 

There are two corrections to the Lenz energy at small densities.  
Including these, the kaon energy $\omega$ obtains the form \cite{Weise,kaon}:  
\begin{equation}
\omega = m_K + \left(\frac{2 m_K}{m_K + \omega} \right) \left( \frac{1} 
{1-a_{K^-n} \xi \rho} \right) \frac{2\pi}{m_R}\, a_{K^-n}\,\rho \ , 
\label{Lenzpl} 
\end{equation} 
where $\xi \rho$ is the inverse correlation length.  The first 
correction factor is a relativistic effect obtained from the 
Klein-Gordon equation, while the second factor is from 
the theory developed by Ericson and Ericson for propagation of 
mesons in nuclear matter.  The relativistic correction decreases 
the kaon energy, while the correlation correction increases it. 

As the density increases further, and the interparticle spacing becomes   
of the order of the range of the interaction, the kaon
will simultaneously interact with two or more nucleons and 
the Lenz approximation will break down.  In the high density limit 
the kaon energy deviates from its rest mass 
by the Hartree potential:
\begin{equation}
   \omega_{Hartree} = m_K + \rho \int v_{K^-n}(r)
           d^3r          \,,\label{Hartree}
\end{equation}
where $v_{K^-n}$ is the $K^-n$ interaction potential.  
As shown in Ref.\ \cite{PPT}, the Hartree potential is considerably less
attractive than the Lenz potential and thus $\omega_{Hartree} > \omega_{Lenz}$.  
The transition from the Lenz to Hartree limits has been recently studied 
\cite{kaon} with a variety of methods including exact calculations for 
simple cubic crystal model of neutron matter.
For reasonable interaction range the transition begins at very low 
densities ($< 0.1 \rho_0$), and the Hartree 
limit is essentially reached by $\sim 3 \rho_0$.  
There are no relativistic corrections to the Hartree energy of kaons 
condensed in the state with zero momentum \cite{kaon} provided 
the $K^-n$ interaction is dominated by the Weinberg-Tomozawa 
vector potential. 

The typical resent results for kaon energy in neutron star matter are 
shown in Fig.10.  The top solid line is obtained with a Wigner-Seitz 
\cite{kaon} calculation for pure neutron matter; it is exact in both 
the low and high density limits, and gives 
essentially the Hartree energy at $\rho > 3 \rho_0$.  
The next curve shows the estimated 
Hartree results for NM containing $\sim$ 15 \% protons, 
below that is the $\omega_{Lenz}$ for neutron matter, while 
the lowest curves uses $K^-N$ scattering amplitude calculated in 
matter including relativistic, correlation and proton fraction corrections 
\cite{Weise}. The recent \cite{apr98} estimate for the 
central density of a 2.0 $M_{\odot}$ star is $\sim 5 \rho_0$, and the results 
shown in Fig.10 indicate that kaon condensation is unlikely at densities 
lower than that.  The heaviest stars ($M = 2.2 M_{\odot}$) made up of 
NM are predicted to have central 
densities of $\sim 7 \rho_0$, and the possibility of kaon condensation in 
their cores can not be ruled out.  However, presence of $\pi^-$, or 
$\Sigma^-$ or quark drops will decrease $\mu_e$ as discussed in the 
following subsections and may 
make kaon condensation unlikely in even the most massive stars.

\subsection{Charged Pion Condensation}

Negatively charged $\pi^-$-mesons will condense in matter 
when their chemical potential becomes lower than $\mu_e$, 
as suggested by Migdal \cite{migdal} and by Sawyer and 
Scallapino \cite{ss}.  In the seventies and eighties this 
possibility was studied by many researchers.  Their work 
has been reviewed in Ref. \cite{kun}, and we discuss it 
rather briefly. 

Fig. 10 suggests that, in absence of interactions $\pi^-$ 
with zero momentum will condense in matter at a rather low density of 
$\sim 1.5 \rho_0$ when $\mu_e$ exceeds their rest mass of 
139 MeV.  However, the $\pi^-n$ S-wave interaction is repulsive, 
and raises the energy of zero momentum pions 
sufficiently above the estimated $\mu_e$.  
The recently found \cite{Yamazaki}, deeply bound $\pi^-$-nucleus
atomic states, are influenced by this repulsion. 
The $K^-n$ S-wave interaction, on the other hand 
is attractive, leading to the possibility of kaon condensation 
discussed in the last section. 

The energy of $\pi^-$, having momenta of the order of 2 fm$^{-1}$, 
$\sim$ 400 MeV without interactions, is 
reduced in matter by the $\pi^-n$ P-wave interaction 
due to the coupling of the $\pi^-$ to $p$-$n$ and $N$-$\Delta$ 
particle-hole states.  Their energies and couplings to $\pi^-$ 
are calculated with 
effective forces described with Landau parameters whose density 
dependence is not well established.  With plausible 
values for the Landau parameters 
a second order transition with $\pi^-$ condensation 
is predicted at $\sim 2 \rho_0$; however it is not expected to 
have a large effect on the EOS \cite{kun}.  

The Japan group also predicts a first order transition to an 
interesting phase with both $\pi^-$ and $\pi^0$ condensation at a 
density of $\sim 4.5 \rho_0$.  It has a significant 
effect on the EOS.  Matter in this phase has spin 
aligned layers as discussed earlier in sec.~3.6 and illustrated 
in Fig. 11.  The condensed $\pi^0$-mesons have momenta in the Z-direction, 
perpendicular to the layers, while that of 
$\pi^-$ is in the X-direction in the plane of the layers.  This 
way, when a proton absorbs a $\pi^-$ form the condensate and becomes 
a neutron its spin direction also gets flipped maintaining the 
attractive interaction with the $\pi^0$ field.  It is a challenge 
to calculate the energy of matter in this interesting  
phase from bare nuclear forces.

The Illinois group, working with bare nuclear forces, predicts the
first order $\pi^0$-condensation at a much lower density of $\sim 1.5
\rho_0$ (sec.~3.6).  However, the total decrease in energy of matter
at $\sim 5 \rho_0$ due to pion condensation, estimated by the two
groups: $\sim$ 80 by the Japan and $\sim$ 60 MeV/nucleon by Illinois,
are not too different.  The recent EOS of the Illinois group
\cite{apr98} contains this energy gain.

\subsection{Hyperonic Matter}

The possibility of hyperons contributing  
to the ground state of dense matter has been considered since 
1959 \cite{camsol,vrpsol74}.  The negatively charged $\Sigma^-$ 
and $\Delta^-$ will occur when their chemical 
potential becomes less than $\mu_n+\mu_e$ in matter.  With 
A18+$\delta v$+UIX$^*$ (A18+$\delta v$) interactions, 
the $\mu_n+\mu_e$ exceeds the rest mass (1197 MeV) 
of $\Sigma^-$ hyperon at density of $\sim$ 2.2 (2.6) $\rho_0$.  If 
the $\Sigma^-$-nucleon interactions are negligible, they 
will occur in matter via the weak interaction 
$e^-+n \rightarrow \Sigma^- +\nu_e$ at these densities
mainly on account of the large $\mu_e \sim 170$ MeV. 
Similarly the neutral hyperons such as the $\Lambda$ 
will occur when their chemical potentials become less than 
$\mu_n$.  For the above interaction models the 
$\mu_n$ exceeds the $\Lambda$ rest mass (1116 MeV) at 
densities of 2.9 and 3.7 $\rho_0$, while the predicted central 
densities of 1.4 $M_{\odot}$ stars are 3.4 and 5.1 $\rho_0$. 

Much less data exist on hyperon-nucleon (YN) interactions than on the
NN, and therefore their models are less constrained.  LOBHF
calculations \cite{bbs98} using the older Nijmegen soft core YN
\cite{mrs89} and either Paris or Argonne $v_{14}$ NN interactions show
that the thresholds for $\Sigma^-$ and $\Lambda$ to appear are not
much moved by YN interactions.  For example, with the Paris NN
interaction the threshold densities with (without) YN interactions are
3.0 (2.7) and 3.6 (3.7) $\rho_0$ for $\Sigma^-$ and $\Lambda$
respectively.  The Nijmegen group \cite{rsy99,rsup} has recently
constructed boson exchange interaction models based on YN and NN data
base using SU(3) symmetry.  LOBHF calculations \cite{vpreh} using
these models for NN as well as YN interactions give 2.2 $\rho_0$ for
the threshold density of $\Sigma^-$, however for $\Lambda$ it is
pushed beyond 7 $\rho_0$.

The above LOBHF calculations use only two-body interactions without boost 
corrections and TNI, and fail to explain the saturation density of SNM 
as mentioned in sec.~3.4.  Both the BHF groups \cite{bbs98,vpreh} 
find that the threshold densities are lowered after including TNI 
effects.  For example, the Catania group predicts them to be 
$\sim$ 2.1 and 2.4 $\rho_0$ with models including TNI and also with 
Dirac-Brueckner calculations.  However, $\Sigma^-NN$ and $\Lambda NN$ 
three body forces should also be included along with the TNI for 
consistency. The binding energies of $\Lambda$-hypernuclei \cite{upu} 
suggest that the $\Lambda NN$ interaction is as strong as the TNI, 
while there is no data on the $\Sigma^-NN$. 

Ignoring three-body forces 
and boost interactions in both nucleon and hyperon matter, 
the LOBHF calculations with the unified NN, YN and YY interaction 
model \cite{vpreh} indicate that admixtures of $\Sigma^-$ lower the NM 
energy by only $\sim 25$ MeV per nucleon at 5 $\rho_0$.  At this density 
the A18+$\delta v$+UIX$^*$ model gives an energy of $\sim 200$ 
for $\beta$-stable NM.  If the energy gain due 
to $\Sigma^-$, is not much changed by three-body interactions, we
should expect $\sim$ 10 \% effects on the EOS of neutron star 
matter at $\rho > 2 \rho_0$ due to hyperons; however, if present 
they would lower the $\mu_e$ and increase the proton fraction $x_p$ 
significantly.  

\subsection{Quark Matter} 
 
When matter is compressed to densities so high that the quark cores of
nucleons overlap substantially, one expects the nucleons to merge and
undergo a phase transition to quark matter (QM).  The EOS of both HM
and QM are necessary to calculate effects of this transition in
neutron stars.  The interface properties are also needed to study the
important mixed phase regions.

At present, lattice QCD can only treat the case of
zero baryon chemical potential and is therefore not useful for
neutron stars.  Lacking a full theory,  
the simple Bag model is used to estimate the QM EOS.  
In this model the QM energy has a volume term denoted by the 
bag constant $B$; it represents the difference in the energies 
of the vacua occupied by hadron and QM, and is responsible for 
the confinement of quarks within nucleons in nuclei. 
The term dominating at high densities is the energy of 
noninteracting u, d and s-quarks; it is calculated neglecting 
the mass of u, and d-quarks, and using a typical value of 
$\sim$ 150 MeV for that of the s-quark.  Since the quark Fermi 
energies are much larger than their masses, the QM properties are 
not too sensitive to the chosen mass for the s-quark. 
There is no one-gluon exchange interaction energy
between quarks of different flavor, while that between quarks of
of the same flavor {\em i} is given by
$(2 \alpha/3 \pi) E_i$ per quark of flavor {\em i} \cite{BC76}. 
Here $E_i$ is the average
kinetic energy per quark, and $\alpha$ is the strong interaction
coupling constant, assumed to have a value of $\sim$ 0.5.
All higher order gluon-exchange interactions are neglected; 
their contribution is presumably subsumed in the bag constant 
whose value is poorly known. 
Two representative values are 
$B=122$ \cite{CLS86} and 200 MeV fm$^{-3}$ \cite{Sat82}.

The equilibrium conditions for uniform QM containing u, d, s-quarks, 
electrons and muons are:
\beq
\mu_u + \mu_e =  \mu_d = \mu_s,  \qquad \mu_{\mu} = \mu_e.
\eeq
The energy densities of charge neutral QM and
NM are compared in Fig.12.
In the interesting region of $\rho \sim 1$ fm$^{-3}$ the total energy
density of quark matter is about 1200 MeV fm$^{-3}$, of which only
122 or 200 MeV fm$^{-3}$ comes from the bag.
Assuming the A18+$\delta v$+UIX$^*$ EOS for NM the first order phase 
transition to QM is indicated at $\sim$ 1 baryon/fm$^3$.  

It has been recently suggested that at high densities QM may have color
superconductivity resulting from non-perturbative
attraction between quarks.  In QM with only u and d-quarks this invariably
leads to the possibility of a diquark condensate which breaks global 
color invariance \cite{RW}. The associated color gap is estimated to be 
of the order of 100 MeV.  
When the s-quarks are included there are many possible phases \cite{SW98}.

\section{Mixtures of Phases in Dense Matter}

The phases of matter considered in the past sections are uniform 
and locally charge neutral, whereas bulk matter needs only to be charge 
neutral on average.  For example, iron metal has positively charged
regions occupied by iron nuclei, and the space in between is negatively charged 
by electrons.  Generally the ground state of matter can have a mixture 
of regions occupied by different phases with the constrain of overall charge 
neutrality. 

Matter in the outer crust of neutron stars, at $\rho \leq 0.002
\rho_0$, is believed to be like terrestrial matter made up of neutron
rich nuclei in electron gas.  The inner crust, on the other hand, is
believed to have a mixture of regions with positively charged NM
composed of neutrons and protons and PNM with only neutrons
\cite{prannr}, both immersed in a nearly uniform electron gas. These
mixtures occur in the density range of $\sim$ 0.002 to 0.6 $\rho_0$,
beyond which uniform NM is believed to be the ground state.  More
recently mixed phases of NM and QM \cite{Gle92,HPS93}, and condensates
of pions, kaons \cite{Schaffner}, or hyperons in NM have also been
considered.

The Coulomb energy of a single phase uniform matter, 
due to fluctuations in the electron and hadron or quark densities, 
is negligible; however that of matter with mixed phases is not.  
For example, matter at $\rho = 0.3 \rho_0$ has drops of NM in 
PNM.  Their size is determined by a competition between
Coulomb and surface energies; large drops have too much
Coulomb energy and small drops to much surface energy per nucleon.
It is necessary to know the energy of the interface to predict 
the nature of the mixed phase region quantitatively.  Those for 
the interface between NM and PNM have been calculated from 
energy-density functionals of NM \cite{prannr}; while those 
for interfaces of NM and QM are not well estimated.  In this 
section we discuss QM and NM mixed phases, to avoid 
duplicating the review by Pethick and Ravenhall \cite{prannr}, 
and because they can influence the mass limit of neutron 
stars. 

\subsection{Equilibrium Conditions for Coexistence of QM and NM}

Neglecting surface and Coulomb effects,  
the equilibrium conditions for the coexistence of 
of QM and NM at zero temperature are that they have equal pressures, and it 
costs no energy to convert a neutron or a proton in NM into quarks in
QM. The last condition amounts to
\beq
  \mu_n=2\mu_d+\mu_u \quad {\rm and} \quad \mu_p=\mu_d+2\mu_u \,. 
\eeq
The electron density, and hence the $\mu_e$, is assumed to be the same
in QM and NM, therefore the $\beta$-equilibrium conditions are:
\beq
   \mu_n=\mu_p + \mu_e  \quad {\rm and} \quad \mu_d=\mu_s=\mu_u+\mu_e \,.
\eeq
And finally, the charge neutrality condition is given by:
\beqa
   f\rho^{ch}_{QM} + (1-f)\rho^{ch}_{NM} = e (\rho_e+\rho_\mu) \,.
\eeqa
Here $\rho^{ch}_{QM,NM}$ are the charge densities of QM and NM, 
$\rho_{e,\mu}$ are the electron and muon densities, and $f$ is the 
fraction of space filled by QM.  A graphical representation of 
these equilibrium conditions is given in ref. \cite{apr98}. 

As was first pointed out by Glendenning \cite{Gle92}, the QM and NM 
mixed phases can span a wide density region.  For example, with the 
A18+$\delta v$+UIX$^*$ NM EOS, the mixed phases begin at 4.6 (3.4) and end at 
11.3 (9.1) $\rho_0$ for $B=200$ (122) MeV fm$^{-3}$ (Fig.12).  At lower densities 
we have uniform NM and QM at higher.  The QM fractional volume increases 
almost linearly from 0 to 1 in this transition range over 
which $\mu_e \rightarrow 0$.  The uniform QM has almost equal number 
of u, d, and s-quarks, and is essentially charge neutral without electrons.
An important consequence of $\beta$-equilibrium is that the QM is {\it
negatively charged} at the beginning of the transition, where  $\mu_e$ 
is large.  By immersing negatively charged drops of QM in the positively
charged NM we can remove some of the high energy electrons 
and increase the proton fraction in NM.  Both of these lower the 
energy density of matter.  

\subsection{Structure of Mixed Phase Matter}

Matter with mixed phases has additional structure due to interfaces
dividing the regions occupied by the two phases.  The surface and
Coulomb energies associated with these interfaces, neglected in above
section, raise the energy of the mixed phase matter as well as
determine the topology and structural length scales.  The surface
energy can presumably be estimated from the surface tension, $\sigma$.
For NM $\sigma\simeq 1$MeV/fm$^2$ whereas for QM it poorly known;
typical values are in the range 10-100 MeV/fm$^2$ \cite{HPS93}).

Denoting the dimensionality of the structures by $D$ ($D=3$ for
droplets and bubbles, $D=2$ for rods and $D=1$ for plates) the surface
energies are generally \cite{prannr}
\begin{equation}  
    {\cal E}_S= D\sigma \frac{4\pi}{3} R^2  \, , \label{ES}
\end{equation}
and, for $D=3$ the Coulomb energy, ${\cal E}_C=(3/5)Z^2e^2/R$,  
where $R$ the size of the structure and
\begin{equation}
Ze = (\rho^{ch}_{QM}-\rho^{ch}_{NM}) \frac{4\pi}{3}\ R^3 \ ,
\end{equation}
is the excess charge of the droplet compared with the
surrounding medium.  General equations for ${\cal E}_C(R,D)$ 
are given in ref. \cite{prannr}. 
Minimizing the energy density
with respect to $R$ we obtain the usual result 
that ${\cal E}_S=2{\cal E}_C$ at equilibrium.  Minimizing with respect to the
continuous dimensionality as well thus determines both $R$ and $D$.
For droplets ($D=3$) the equilibrium radius is found to be:
\begin{equation}
    R\simeq 4.0\, {\rm fm} \, \left(\frac{\sigma}{1{\rm MeV/fm}^2}
\right)^{1/3} \left( 
\frac{\rho^{ch}_{QM}-\rho^{ch}_{NM}}{e\rho_0/2}\right)^{-2/3}.  \label{R}
\end{equation}
A droplet of
symmetric NM in vacuum has a surface tension $\sigma = 1$
MeV$\cdot$fm$^{-2}$ for which (\ref{R}) gives $R\simeq$ 4 fm, which
agrees with the fact that nuclei like $^{56}$Fe are the most stable form
of matter at low density. For QM droplets both the surface tension and
charge densities are larger but the estimates of $R$ are similar.

At the beginning of the mixed phase region we expect that spherical 
droplets of QM, with $R\sim 5$ fm, will form a BCC lattice in 
uniform NM.  They would have baryon number of the order of few 
hundred, and a negative charge of similar magnitude.  As density 
increases and $f$ approaches 0.5, the drops would merge and form 
rods, which merge further on to form sheats.  When $f > 0.5$
the NM sheats break up into rods, and then into drops and eventually 
disappear when $f=1$.  This scenario is similar to that in the 
inner crust; at low densities there are drops of NM occupying a small 
fraction of space.  By $\rho \sim 0.6 \rho_0$ NM occupies all space 
via a similar set of mixed phases. 

An other effect of the Coulomb and surface energies is that they 
decrease the density range covered by the mixed phase region. 
In particular, the lower density edge of this interesting region may 
be pushed up by almost $\rho_0$ if $\sigma$ is in the 10 to 50 MeV 
fm$^{-2}$ range \cite{HPS93}.  The energy density of the mixed phase
matter is also raised by a few MeV fm$^{-3}$ in this case.  Finally, 
if $\sigma$ were to be large 
$( \ga 70 $ MeV fm$^{-2})$ the
mixed phases may not be energetically favorable, and there will 
be a simple first order phase transition from NM to QM with a 
density discontinuity. 
One should bear in mind that even if the droplet phase
were favored energetically, it may not be realized in practice if
the time required to nucleate QM drops is too long compared to 
pulsar ages.

\section{Neutron Star Observations and Predictions}

The gross structure of neutron stars has been predicted using very many 
EOS, phenomenological as well as based on realistic models of nuclear forces
\cite{glenb,Lattimer}.  Of these we consider only those based on realistic 
models primarily because one can always find phenomenological energy density 
functionals or Lagrangians which reproduce their EOS.  

Typical results for nonrotating stars with maximum mass, and with
$M=1.4 M_{\odot}$, obtained by recent calculations, are listed in
Table 2.  The results for A18 without boost correction $\delta v$ are
listed primarily for reference.  This correction is unambiguous
\cite{FPF95}, and must be added to obtain reliable results.  Those for
A18+$\delta v$ are also to be taken less seriously, because it gives
too large value for $\rho_0$.  The TNI used with the Paris NN
interaction $\cite{baldo97}$ is of the Urbana form with parameters
determined by reproducing the empirical SNM properties (see sec.~3.4).
In A18+UIX and Paris+TNI models the $\delta v$ is not considered
explicitly; it is approximately subsumed in the TNI fitted to data.
Of the three Bonn models, Bonn A comes closest to reproduce the
empirical properties of SNM \cite{bromac} with Dirac-Brueckner (DBHF)
method.  These calculations include the $\delta v$ as well as
many-body forces generated via $Z$-diagrams.

\begin{table}
\caption{Properties of maximum mass $(M_m)$ and 1.4 $M_{\odot}$ neutron stars in 
        $M_{\odot}$, $\rho_0$ and km.}
\vspace{0.2cm}
\begin{tabular}{lrrccccc}
\hline
Interactions & Calc. & Ref. & $M_m$ & $\rho_c(M_m)$ & $R(M_m)$ & 
$\rho_c(1.4)$ & $R(1.4)$ \\
\hline
A18 & Var. & \cite{apr98} & 1.67 & 11.1 & 8.1 & 7.0 & 8.2 \\
A18+$\delta v$ & Var. & \cite{apr98} & 1.80 & 9.4 & 8.7 & 5.1 & 10.1 \\
A18+$\delta v$+UIX$^*$ & Var. & \cite{apr98} & 2.20 & 7.2 & 10.1 & 3.4 & 11.5 \\
A18+UIX & Var. & \cite{apr98} & 2.38 & 6.0 & 10.8 & 2.9 & 12.1 \\
Paris+TNI & BHF & \cite{baldo97} & 1.94 & 8.3 & 9.5 & 4.0 & 11.1 \\
Bonn A & DBHF & \cite{baldo97} & 2.10 & 6.7 & 10.6 & 3.1 & 11.7 \\
\hline
\end{tabular}
\label{tb:nstarmr}
\end{table}

The A18+$\delta v$+UIX$^*$, A18+UIX, Paris+TNI and Bonn-A (DBHF) models 
come close to reproducing the empirical $\rho_0$; the later two fit 
the SNM binding energy; while the former 
models fit binding energies of light nuclei via exact calculations, 
since the energy of SNM can not yet be calculated reliably. 
Nevertheless these four ``{\em realistic}'' models of NM 
give rather similar results which are not too different from those 
of the 1988 calculations of Wiringa, Fiks and Fabrocini \cite{WFF} 
with the older Urbana-Argonne interactions now replaced with A18 and UIX.  

The effect of the possible appearance of QM drops in high density matter 
has been studied with the A18+$\delta v$+UIX$^*$ model.  The $M_m$ is 
reduced to 2.02 and 1.91 $M_{\odot}$ for bag-constant values $B=$ 200 
and 122 MeV Fm$^{-3}$ respectively, while the predictions for 1.4 
$M_{\odot}$ stars remain unchanged.  Presence of either kaons or 
hyperons in dense matter is unlikely to have much of an effect on the 
1.4 $M_{\odot}$ stars due to their low central density, while that on the 
mass limit is difficult to estimate quantitatively.  For example, if kaons were 
to condense in matter at $\rho = 5 \rho_0$ and limit $\rho_c$ to $<$ 
5 $\rho_0$, the $M_m$ of the four realistic models will drop to 
$\sim$ 2.0, 2.3, 1.7 and 2.0 $M_{\odot}$ respectively; while if hyperons 
were to lower the energy of matter at $\rho=5 \rho_0$ by 25 MeV per 
baryon, the $M_m$ would be reduced by $\sim$ 0.2 $M_{\odot}$. 

The mass radius relation obtained with models based on the A18 interaction 
are shown in Fig.13.  Results of A18 and A18+$\delta v$ models are given 
primerily for comparison. 
As expected the harder EOS give larger $M_m$ and predict larger radii. 
The differences between the radii predicted by the realistic models  
is only $\sim$ 10 \%. 

\subsection{The Mass Limit}

The observed mass of Hulse-Taylor pulsar B1913+16 of 1.4411 $\pm$ 
0.00035 \cite{Thor99} shows that $M_m > 1.44 M_{\odot}$.  
All the radio pulsars in known neutron star, and neutron star-white dwarf 
binaries have masses with lower limits less than $1.44 M_{\odot}$.
The X-ray pulsar Vela X-1, which orbits a supergiant, however 
is consistently estimated to have a larger mass of $\sim 1.9$. 
The motion of this star is perturbed from being pure Keplerian, 
presumably by tidal forces exerted by the neutron star, and its  
present mass estimate, $1.87^{+0.23}_{-0.17}$
\cite{Barziv}, indicates that $M_m > 1.7 M_{\odot}$  
at 95 \% confidance level. 
Finally, if the QPO's indeed originate from the innermost 
stable orbit \cite{zss97,miller}, then $M_m > 2 M_{\odot}$. 
These mass limits are compatible with predictions of realistic NM 
models.

On the other hand there is no evidence that SN 1987A produced a
neutron star.  Its observed luminosity is now well below the 10$^{38}$
ergs/s Eddington limit, suggesting that no neutron star was produced in
this supernovae \cite{cheva}.  If we assume that the total mass,
$M_{Tot}$, of the collapsed core plus the matter that fell back on to
the core after the explosion, went into a black hole, then neutron
star $M_m$ must be less than $\sim 0.9 M_{Tot}$.  The factor 0.9 takes
into account the $\sim$ 10 \% gravitational binding energy of the
neutron star.  Bethe and Brown \cite{Bebhole} estimate $M_{Tot} \sim
1.73 M_{\odot}$ using supernovae calculations by Wilson and Mayle, and
conclude that $M_m < 1.56 M_{\odot}$.  Uncertainties in these
arguments have been discussed by Zampieri {\em et. al.} \cite{shapi}.
If the conclusion is found to be valid, then there must be other
explanations for the Vela X-1 observations and the QPO, and the NM
prediction for the $M_m$ is too large.

\subsection{Temperatures, Cooling and Radii}

Neutron stars are born with interior temperatures of the order
$10^{12}$~K, but cool rapidly via neutrino emission to temperatures of
the order $10^{10}$~K within minutes and $\la10^6$~K in $10^5$ yr.
Spectra observed in X-ray or UV bands for nearby pulsars have in some
cases black-body components from which surface temperatures of order
$T\sim 10^6$~K are extracted for pulsars of age $10^3-10^6$~years.  It
is, however, unclear how much of the observed radiation is due to
pulsar phenomena, to a synchrotron-emitting nebula or to the neutron
star itself.  In other cases upper limits have been set from the
absence of X-rays.  The surface temperatures are compatible with
predictions from standard modified URCA cooling processes
\cite{Maxwell}
\beq
 n+n \to n+p+e^-+\bar{\nu}_e ,\quad n+p+e^-\to n+n+\nu_e .
\eeq
Faster cooling processes as direct URCA or due to quark matter, kaon
or hyperon condensates generally lead to considerably lower
temperatures \cite{Page}. To be consistent with observed surface
temperatures the exotic coolant can only exist in a minor portion of
the neutron star or it is superfluid whereby cooling is suppressed
by factors of $\propto\exp(-\Delta/T)$, where $\Delta$ is the pairing gap.

The Hubble Space Telescope (HST) has observed one
thermally radiating neutron star RX J185635-3754 with surface
temperature $T\simeq 6\times 10^5$~K$\simeq 50$~eV \cite{Walter}. Its
distance is less than 120 pc from Earth and should soon be determined
more accurately by HST parallax measurements. Circumstancial evidence
indicate a distance of $\sim$80~pc \cite{Lattimer} which leads to a
black-body radius of $\sim 12-13$~km from its luminosity and
temperature.  Such radii would agree well with predictions of
realistic NM EOS (Fig.13) for $M\simeq 1-2M_\odot$.

\subsection{Glitches and Superfluidity}

Sudden spin jumps, called glitches, superimposed upon otherwise
gradual spin down have been observed in most of the younger isolated
pulsars \cite{Epstein}.  Since their discovery the Crab and Vela
pulsars have each produced about a dozen glitches with period changes
$\Delta P/P$ of the order of $10^{-8}$ and $10^{-6}$ respectively.  In
post-glitch relaxation most of the period increase $\Delta P$ decays.

Many mechanisms have been proposed to explain the glitches \cite{gaps}.  The most 
plausible of these attributes glitches to the angular momentum stored 
in the rotating superfluid neutrons in the inner crust
\cite{Pines,prannr}.  The magnetic torque slows down the crust and most of
the star except for these superfluid neutrons. Their angular momentum 
is stored in vortices pinned to nuclei in the inner core, until an instability
occurs that leads to vortex depinning and sudden angular momentum transfer to
the crust, leading to the glitch. 
At subnuclear densities in the crust, $^1S_0$ pairing
between neutrons leads to gaps of order $\sim$1~MeV \cite{gaps}.
In NM at $\rho > \rho_0$  
this pairing gap vanishes, but $^3P_2$ pairing of neutrons and $^1S_0$
pairing of protons may occur \cite{gaps,baldo98}. 

The size of the glitches sets a lower limit on the moment of inertia
of the superfluid in the inner crust which in turn sets a lower limit
on the neutron star radius for a given mass \cite{Epstein}. Assuming 
that the mass of Vela pulsar is $1.4M_\odot$, a conservative limit on its
radius is $R\ge 9$~km; it is compatible with predictions of
most EOS.

\section{Conclusions}

Since the discovery of pulsars a significant effort has been devoted 
to accurately calculate properties of dense NM from realistic models of 
nuclear forces.  Exact calculations of NM are still out or reach, 
however the new AFDMC methods (sect. 3.7) may eventually succeed.  The 
present variational upper bounds seem to be above the true energies 
by $\sim$ 12 \%.  Such an error does not have serious consequences on the 
predicted properties of neutron stars.  For example, an EOS obtained 
by reducing the variational energies, without rest mass terms, by 
12 \% reduces the maximum mass of A18+$\delta v$+UIX$^*$ model by 2.3 \% 
to 2.14 $M_{\odot}$, 
and the radius of 1.4 $M_{\odot}$ star 
by 2.9 \% to 11.2 km.  Larger uncertainties stem from the fact that 
the double $\pi^0$ and $\pi^-$ condensation scenario illustrated in Fig.11 
has not yet been calculated with realistic interactions, though it appears 
unlikely that it will influence the NM EOS by much more than 10 \%. 

Local models of two-nucleon interaction seem to be now converging.
The predictions based on the 1988 calculations with Argonne 14
interaction are not too different from those of the 1998 calculations
with the more accurate A18.  It also seems likely that the local
models give a fairly accurate description of two-nucleon interaction.
A concern is that the present models of TNI are based on fits to a
rather limited set of data, and are not as precise as the
NN-interaction models.  However, addition of the UIX$^*$ TNI to the
A18+$\delta v$ increases the maximum mass by $\sim$ 20 \% and $R(1.4)$
by 13 \% (Table 2).  These changes may be important but they are not
very large.

The present models of kaon-nucleon and hyperon-nucleon interactions 
are based on very limited data, and we have none on $K^-NN$ and 
$\Sigma^-NN$ three-body forces.  These could have significant effect 
on the threshold densities for kaons and hyperons to appear in 
dense matter.  Hopefully advances in QCD and quark-models will provide 
a more rigorous framework to describe these interactions, and calculate 
properties of quark matter.  The bag model estimates of QM EOS may have 
significant corrections at densities of interest in neutron stars.

>From present observations there seem to be three possible scenarios for 
the limiting mass of neutron stars.  If QPO's are indeed due to accretion 
from the innermost stable orbit, then the NM predictions of $M_m \sim 2.2 
M_{\odot}$ are reasonable, and strange baryons and quark drops do not 
soften the EOS of matter at $\rho < 7 \rho_0$ significantly.  If the 
Vela X-1 mass measurement is correct, but QPO's have some other origin, 
then $M_m$ could be $\sim$ 1.8 $M_{\odot}$, indicating some softening of 
the NM EOS.  However, if the present interpretation of QPO's and 
Vela X-1 mass measurements are both faulty, and $M_m$ is as small as 1.56 $M_m$ 
as estimated from the absence of a neutron star in SN 1987A, then a significant 
softening of the NM EOS by phase transitions is 
indicated.  Further observations will hopefully clear this situation. 

Phase transitions such as NM to QM, can soften the EOS 
significantly.  Fortunately these can have a measurable effect on the 
spin down of a rapidly rotating star in favorable cases, 
as has been recently pointed out \cite{gpw97,rot}.  Consider the case 
of a rapidly rotating star whose central density is close to a first 
order phase transition.  As the star slows and the central pressure 
increases due to decrease of the centrifugal force, the core matter 
will change its phase and become more dense at a critical angular 
velocity $\Omega_c$.  This decreases the moment of inertia, which 
assumes the characteristic form: 
\begin{equation}
I = I_0\left( 1 +c_1\Omega^2-c_2(\Omega_c^2-\Omega^2)^{3/2} + ... \right) .
\label{Igen}
\end{equation}
for $\Omega < \Omega_c$. Here, $c_1$ and $c_2$ are small parameters
proportional to the density difference between the two phases, and 
$c_2=0$ for $\Omega>\Omega_c$.

In order to make contact with observation, the temporal behavior of
angular velocities must be considered. The pulsars slow down at a rate
given by the loss of rotational energy, believed to be given by: 
$d(\frac{1}{2}I\Omega^2)/dt\propto -\Omega^{n+1}$, where $n=3$ for
dipole radiation, Eq. (\ref{dE}) 
and $n=5$ for gravitational radiation.  With the
moment of inertia given by Eq. (\ref{Igen}) the angular velocity can
be calculated.  The corresponding braking index,
$n(\Omega)=\ddot{\Omega}\Omega/\dot{\Omega}^2$, depends on the second
derivative of the moment of inertia, $I''=dI/d^2\Omega$. 
Using Eq. (\ref{Igen}) we obtain:
\begin{eqnarray}
n(\Omega)  \simeq n - c_1\Omega^2
+c_2\frac{\Omega^4}{\sqrt{\Omega_c^2-\Omega^2}} \,.\label{n}
\end{eqnarray}
which exhibits a characteristic $(\Omega_c-\Omega)^{-1/2}$
singularity as $\Omega$ approaches $\Omega_c$ from below.
Observations of the braking index of a rapidly rotating, new born 
pulsar would be very interesting.   

All realistic NM EOS predict that the radius of neutron stars with a 
mass of 1 to 1.5 $M_{\odot}$ is $\sim$ 11 to 12 km.  
Future high resolution Chandra and
XMM space observatories will hopefully be able to 
measure black-body spectra and
detect gravitationally redshifted spectral lines from several 
stars.  Such observations will help determine
masses, radii and temperatures uniquely if the distance of the star is known.
It is important to know the radius of a 1.4 $M_{\odot}$ star, because that 
would test the EOS in the $\rho \la 3 \rho_0$ region in which large 
modifications of NM EOS are not expected on the basis of our present, naive 
estimates of kaon-nucleon and $\Sigma^-$-nucleon interactions.  

\section{Acknowledgements}

The authors would like to thank J. Carlson, L. Engvik, S. Fantoni, 
M. Hjorth-Jensen, F. Lamb, S. Pieper, S. Shapiro and R. Wiringa for 
discussions and communications.  This work has been partly supported 
by US National Science Foundation under Grant PHY 98-00978.

\newpage

\begin{figure}
{\centering\mbox{\psfig{file=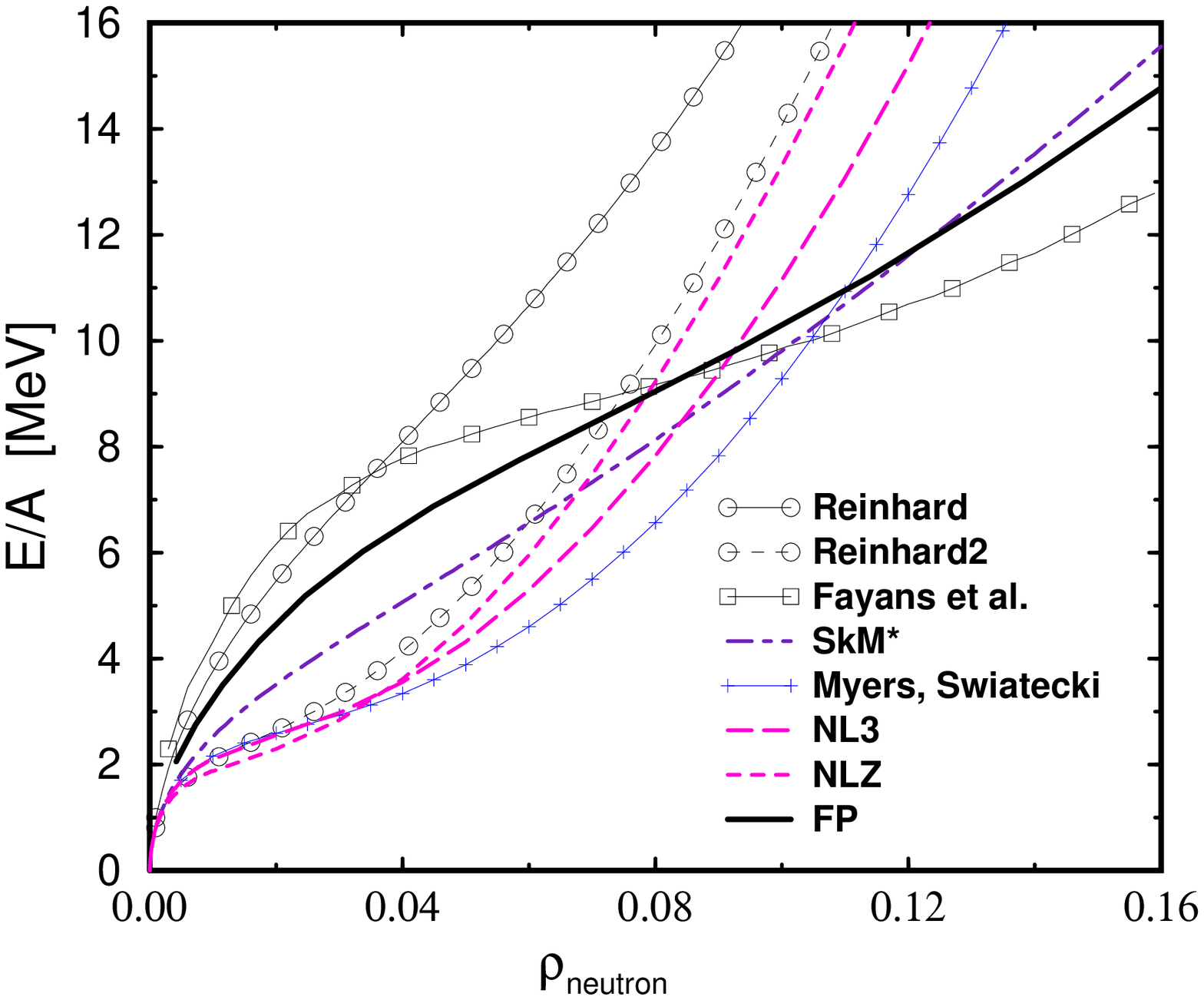,height=10cm,angle=0}}}
\caption{The energy per nucleon in uniform neutron matter 
at low densities. A comparison of
various effective interactions: Reinhard et al. \protect\cite{Reinhard},
Myers et al. \protect\cite{Myers}, Fayans et al. \protect\cite{Fayans}, 
SkM* \protect\cite{SkM*},
Nijmegen NL3 and NLZ \protect\cite{SKR93,SKT94}
with results of calculations using realistic interactions
(FP) \protect\cite{FP}.}
\label{fig1}
\end{figure}

\begin{figure}
{\centering\mbox{\psfig{file=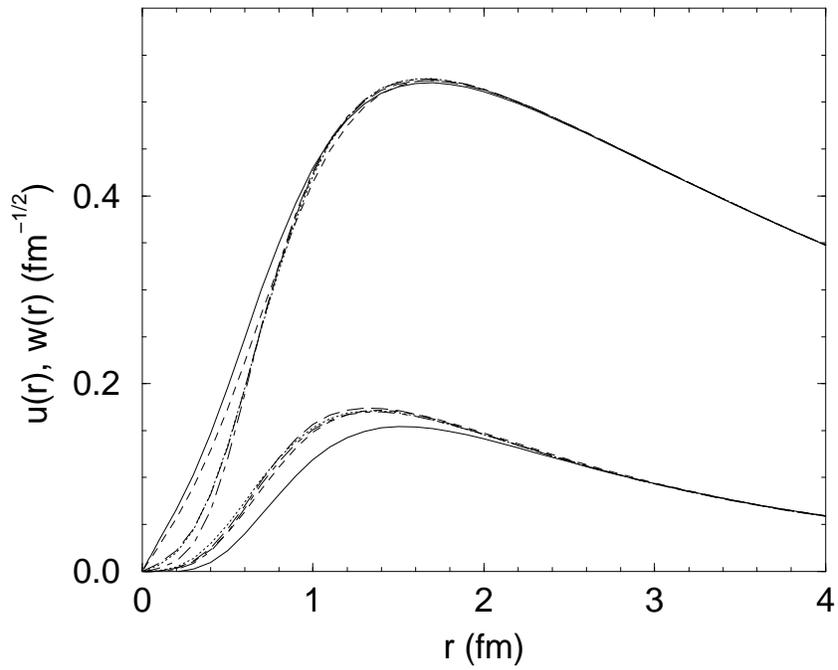,height=10cm,angle=0}}}
\caption{Deuteron radial wave-functions, $^3S_1$ u(r) (upper curves) and 
$^3D_1$ w(r) (lower 
curves) predicted by modern $NN$ interactions: CD-Bonn (solid), 
Nijm-I (dashed), Nijm-II (dash-dotted), Reid 93 (dotted) and A18 
(long-dashed).}
\label{fig2}
\end{figure}

\begin{figure}
\mbox{\psfig{file=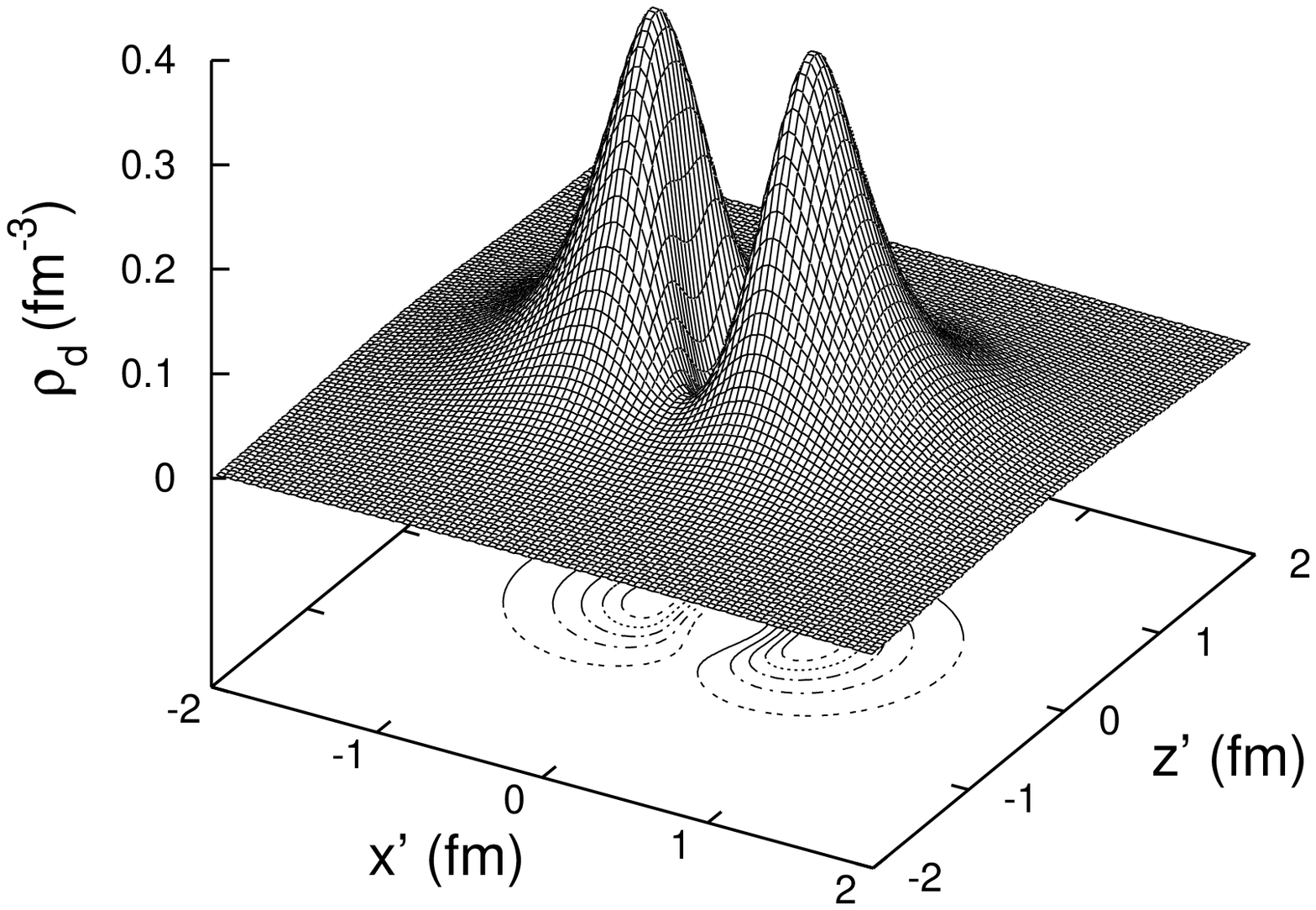,height=15cm,angle=-90}}
\mbox{\hspace{5cm}\psfig{file=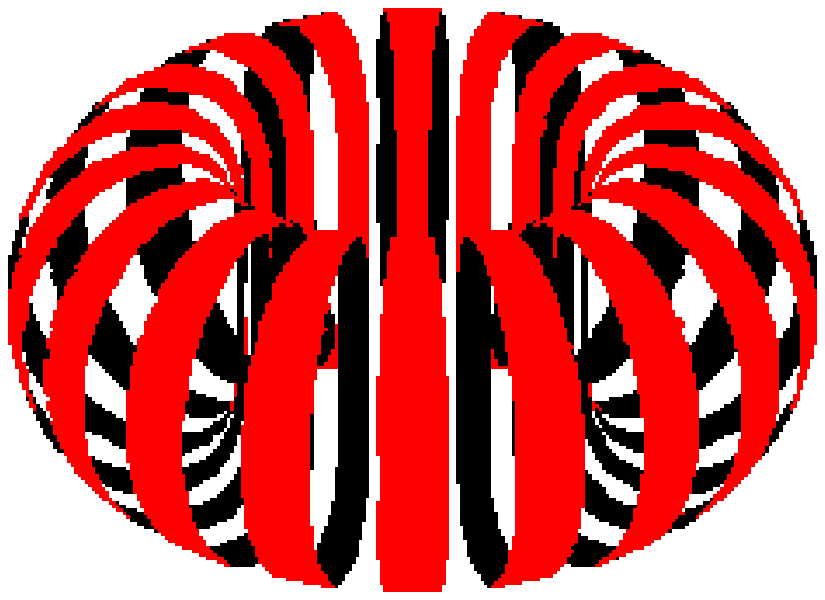,height=4cm,angle=0}}
\caption{The cross section of the density distribution of the 
deuteron in state with spin projection $M=0$ (top part A); 
and the equi-density surface of the deuteron at half maximum density 
(bottom part B).  The toroidal equidensity surface has a diameter 
of $\sim$ 1 fm, and thickness of $\sim$ 0.8 fm.}
\label{fig3}
\end{figure}

\begin{figure}
{\centering\mbox{\psfig{file=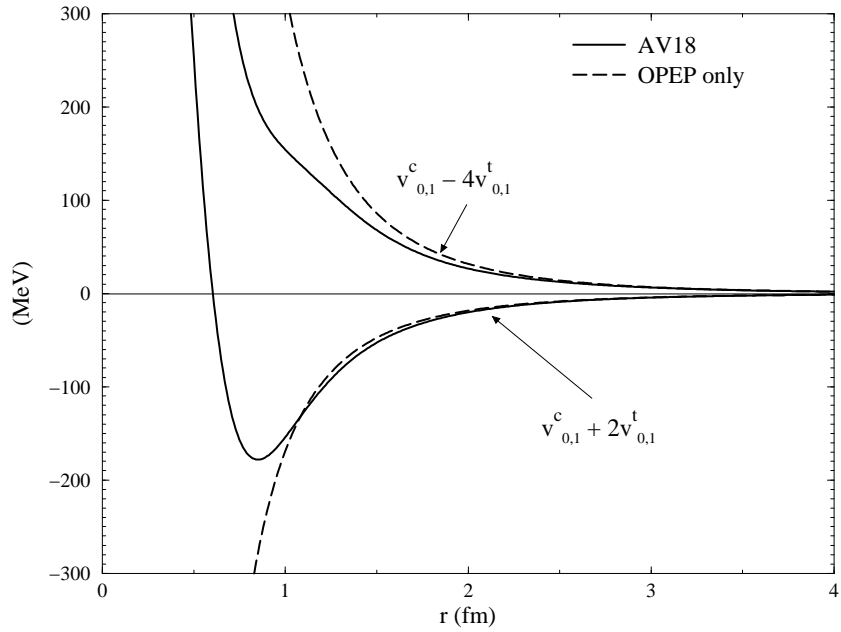,height=10cm,angle=-90}}}
\caption{Static part of the $NN$ potential in the deuteron in spin 
projection $M=0$ state.  The upper curves show the potential along 
the $Z$-axis ($\theta = 0$), while the lower curves show it  
in the $X-Y$ plane.  The $v^{c,t}_{0,1}(r)$ denote the central and 
tensor components of the $NN$ interaction in the deuteron. }
\label{fig4}
\end{figure}

\begin{figure}
{\centering\mbox{\psfig{file=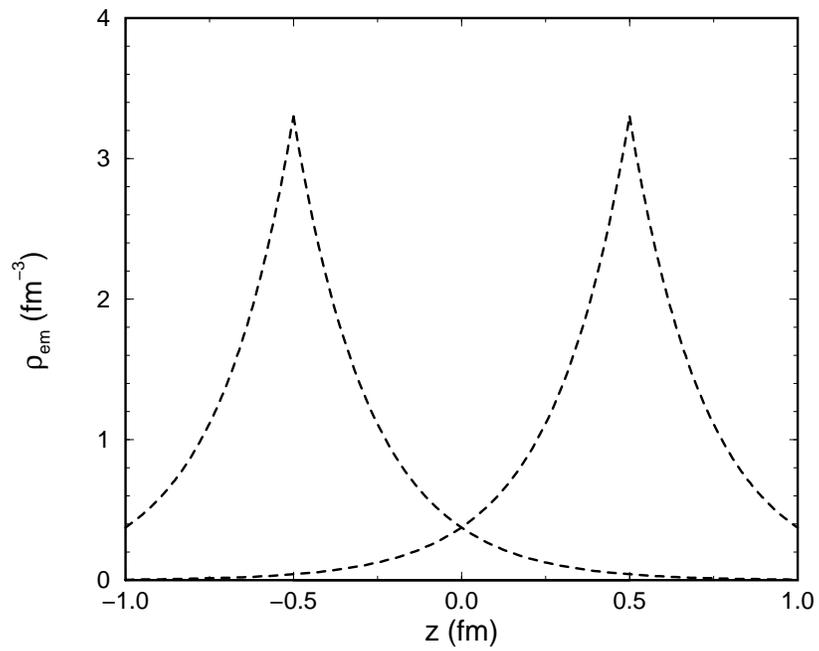,height=10cm,angle=0}}}
\caption{Charge densities of two protons located one fm apart 
at $Z = \pm 0.5 $, obtained by 
inverting the dipole approximation to proton charge form factor.  
The sharp peaks at $Z = \pm 0.5 $ are unphysical, they will 
be rounded off by relativistic corrections and improved data on 
proton form factor.}
\label{fig5}
\end{figure}

\begin{figure}
{\centering\mbox{\psfig{file=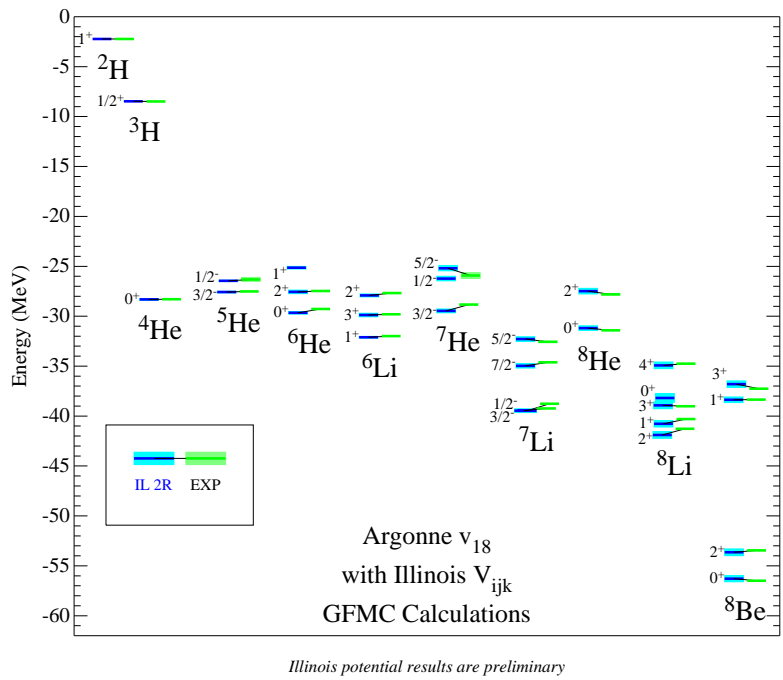,height=10cm,angle=-90}}}
\caption{The observed energies of all bound and quasi-bound states 
of up to eight nucleons are compared with the preliminary results of GFMC 
calculations with $< 2 \%$ errors, using A18 model of $v_{ij}$ and 
Illinois model 2R of $V_{ijk}$ }
\label{fig6}
\end{figure}

\begin{figure}
{\centering\mbox{\psfig{file=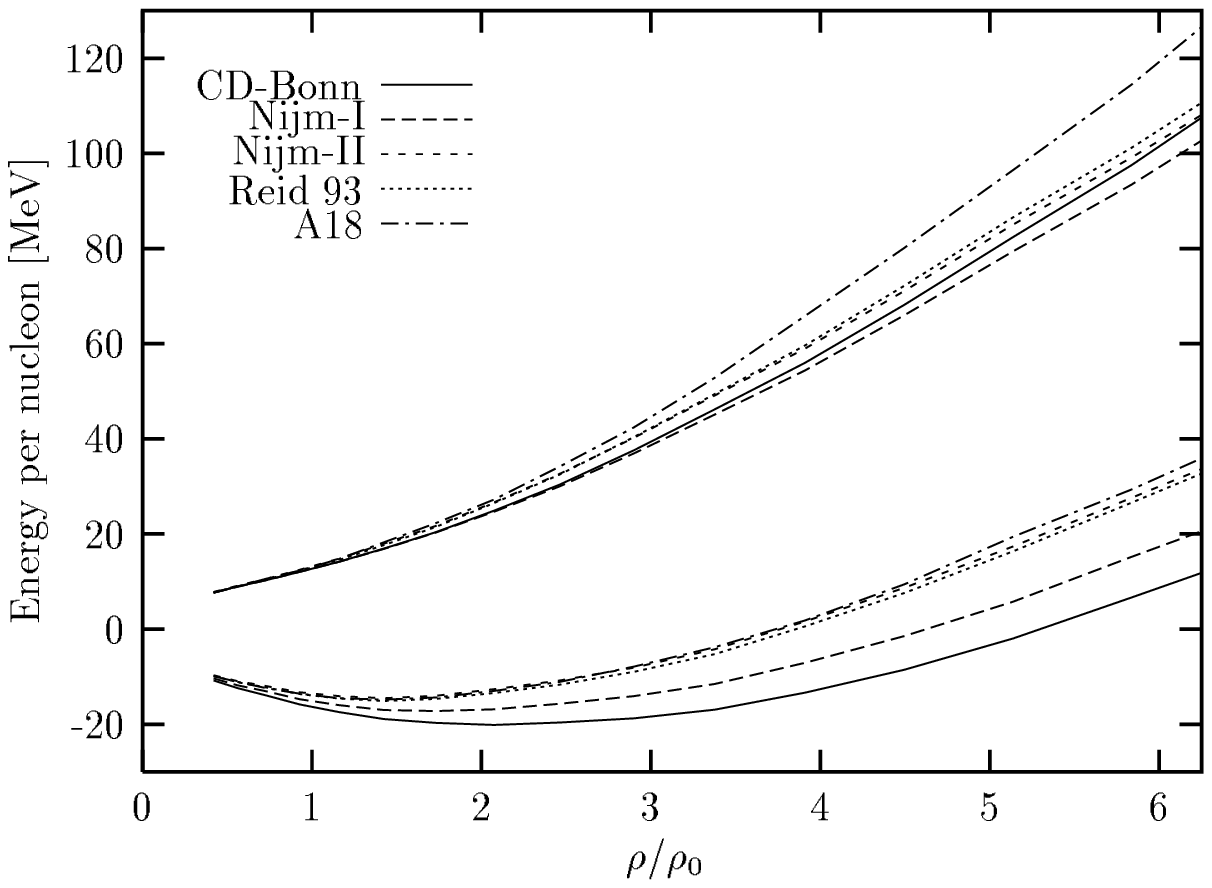,height=25cm,angle=0}}}
\caption{Energy per nucleon of pure neutron matter (upper curves) 
and symmetric nuclear matter (lower curves) 
as function of density, calculated from the five modern models 
of $v_{ij}$ with the  LOBHF method \protect\cite{Eng97}.}
\label{fig7}
\end{figure}

\begin{figure}
{\centering\mbox{\psfig{file=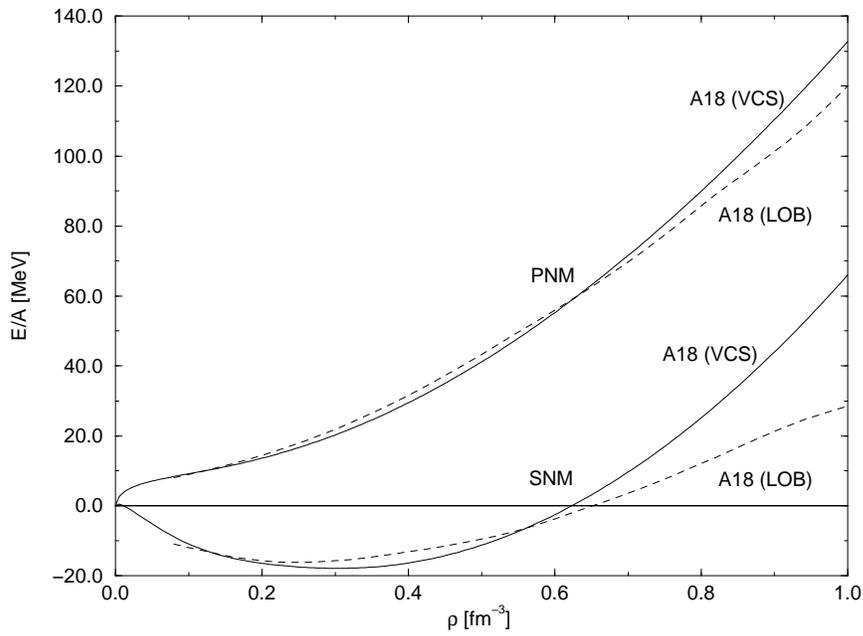,height=10cm,angle=-90}}}
\caption{Comparison of the energies of PNM and SNM obtained for
the A18 model with the variational chain summation (VCS) method 
\protect\cite{apr98} and LOBHF (LOB) method 
\protect\cite{Eng97}.  The true results for this interaction are 
expected to be few MeV below the VCS. }
\label{fig8}
\end{figure}

\begin{figure}
{\centering\mbox{\psfig{file=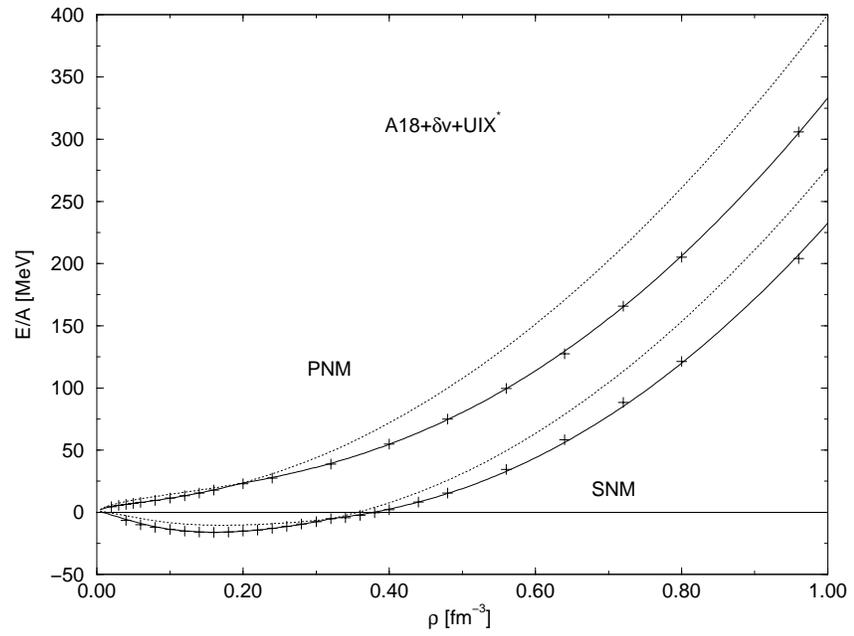,height=10cm,angle=-90}}}
\caption{The PNM and SNM energies for the A18+$\delta v+UIX^*$ model, 
and the fits to them using effective interactions. The full lines
represent the stable phases, and the dotted lines are their extrapolations.
From \protect\cite{apr98}.}
\label{fig9}
\end{figure}

\begin{figure}
{\centering\mbox{\psfig{file=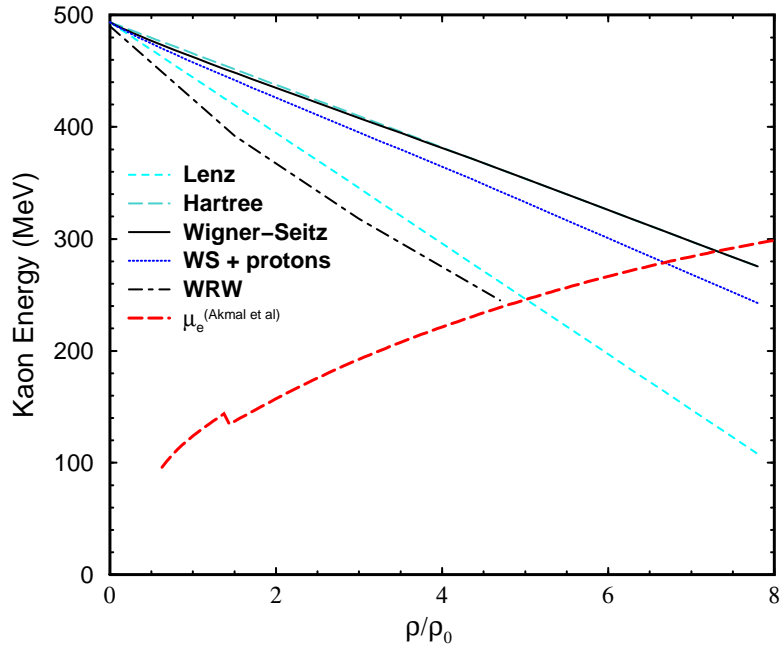,height=10cm,angle=-90}}}
\caption{Kaon energy as function of NM density calculated with 
the various approximations discussed in the text is compared 
with the electron chemical potential $\mu_e$ calculated from 
A18+$\delta v$+UIX$^*$ model in Ref.\ \protect\cite{apr98}.  }
\label{fig10} 
\end{figure}

\begin{figure}
\vspace{1cm}
\mbox{\hspace{2cm}\psfig{file=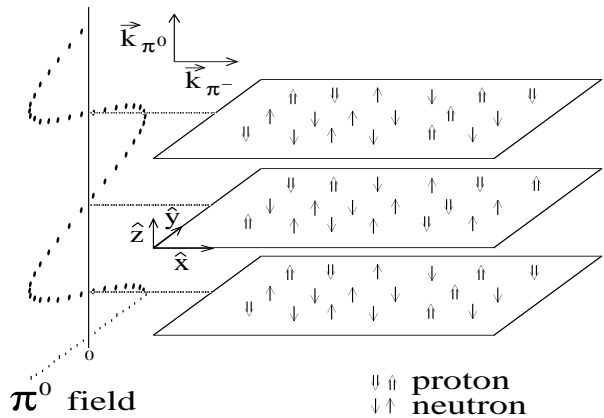,height=8cm,width=8cm,angle=0}}
\vspace{-1cm}
\caption{Schematic drawing of the spin arrangement of neutrons and 
protons in a phase with $\pi^0$ condensation. The nucleons are expected 
to reside mostly in nodal plains of the $\pi^0$ field, where the field 
gradient is largest. The charged $\pi^-$ may condense with momenta 
perpendicular to that of the $\pi^0$ field.}
\label{fig11}
\end{figure}

\begin{figure}
{\centering\mbox{\psfig{file=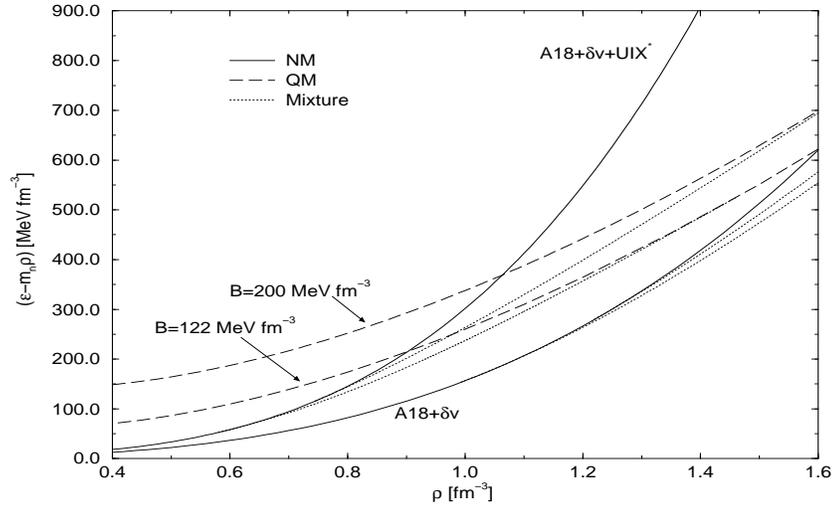,height=8cm,width=8cm,angle=-90}}}
\caption{ The energy densities of NM (full lines), QM (dashed lines) 
and mixtures (dotted lines) from \protect\cite{apr98}. The rest mass 
contribution $m_n \rho$ to the energy density of NM is subtracted from 
the results of all the models for easier comparison. }
\label{fig12}
\end{figure}

\begin{figure}
{\centering\mbox{\psfig{file=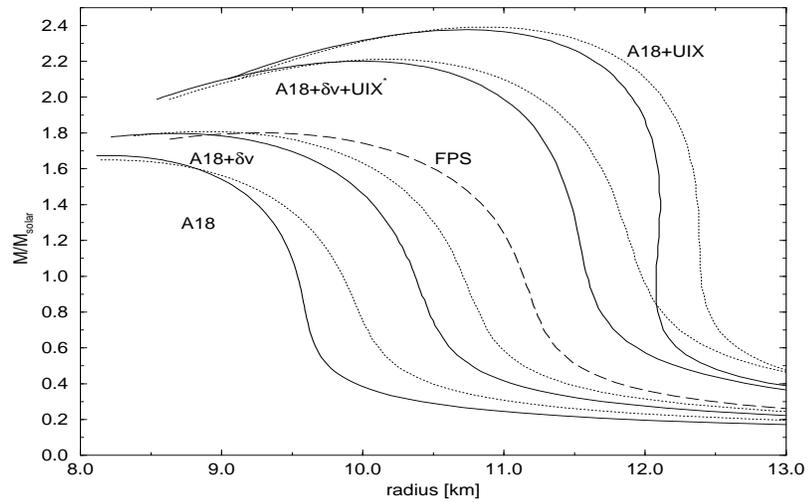,height=8cm,width=8cm,angle=-90}}}
\caption{Neutron star gravitational mass,$M(R)$, in solar masses vs.
radius in kilometers for the four models described in the text.
Full curves are for $\beta$-stable matter and dotted ones for
pure neutron matter. The dashed curve, FPS, is from \protect\cite{FP}.
Fig. from \protect\cite{apr98}.}
\label{fig13}
\end{figure}

\end{document}